\documentclass[a4paper,noarxiv,onecolumn]{quantumarticle}

\usepackage[utf8]{inputenc}
\usepackage[english]{babel}
\usepackage[T1]{fontenc}
\usepackage[ruled,vlined]{algorithm2e}

\usepackage{graphicx}
\usepackage{amsmath}
\usepackage{bbold}
\usepackage{epsfig}
\usepackage{mathtools}

\DeclarePairedDelimiter\ket{\lvert}{\rangle}

\usepackage{hyperref}
\usepackage{breakurl}
\usepackage{listings}
\usepackage{pgfplots}
\usetikzlibrary{patterns,shapes.arrows}
\pgfplotsset{compat=newest}

\begin{document}

\title{High performance Boson Sampling simulation via data-flow engines}

\author{Gregory Morse}
\affiliation{Department of Programming Languages and Compilers, E\"otv\"os  Lor\'and  University, Budapest, Hungary}

\author{Tomasz Rybotycki}
\affiliation{AstroCeNT -- Particle Astrophysics Science and Technology Centre -- International Research Agenda, Nicolaus Copernicus Astronomical Center, Polish Academy of Sciences, Warsaw, Poland}
\affiliation{ACC Cyfronet, University of Science and Technology, Cracow, Poland}
\affiliation{Systems Research Institute, Polish Academy of Sciences, Warsaw, Poland}

\author{Ágoston Kaposi}
\affiliation{Department of Programming Languages and Compilers, E\"otv\"os  Lor\'and  University, Budapest, Hungary}
\affiliation{Wigner Research Centre for Physics of the Hungarian Academy of Sciences, Budapest, Hungary}

\author{Zoltán Kolarovszki}
\affiliation{Department of Programming Languages and Compilers, E\"otv\"os  Lor\'and  University, Budapest, Hungary}
\affiliation{Wigner Research Centre for Physics of the Hungarian Academy of Sciences, Budapest, Hungary}

\author{Uro\v{s} Stoj\v{c}i\'c}
\affiliation{Maxeler Technologies, a Groq company, 16192 Coastal Hwy, Lewes, United States}

\author{Tamás Kozsik}
\affiliation{Department of Programming Languages and Compilers, E\"otv\"os  Lor\'and  University, Budapest, Hungary}

\author{Oskar Mencer}
\affiliation{Maxeler Technologies, a Groq company, 16192 Coastal Hwy, Lewes, United States}

\author{Micha\l\ Oszmaniec}
\affiliation{Center for Theoretical Physics, Polish Academy of Sciences}
\affiliation{NASK National Research Institute, ul. Kolska 12
01-045 Warszawa, Poland}

\author{Zoltán Zimborás}
\affiliation{Wigner Research Centre for Physics of the Hungarian Academy of Sciences, Budapest, Hungary}
\affiliation{Department of Programming Languages and Compilers, E\"otv\"os  Lor\'and  University, Budapest, Hungary}
\affiliation{Algorithmiq Ltd, Kanavakatu 3C 00160 Helsinki, Finland}

\author{Péter Rakyta}
\affiliation{Department of Physics of Complex Systems, E\"otv\"os  Lor\'and  University, Budapest, Hungary}
\affiliation{Wigner Research Centre for Physics of the Hungarian Academy of Sciences, Budapest, Hungary}

\begin{abstract}
In this work, we generalize the Balasubramanian-Bax-Franklin-Glynn (BB/FG) permanent formula to account for row multiplicities during the permanent evaluation and reduce the complexity of permanent evaluation in scenarios where such multiplicities occur.
This is achieved by incorporating n-ary Gray code ordering of the addends during the evaluation. We implemented the designed algorithm on FPGA-based data-flow engines and utilized the developed accessory to speed up boson sampling simulations up to $40$ photons, by drawing samples from a $60$ mode interferometer at an averaged rate of $\sim80$ seconds per sample utilizing $4$ FPGA chips. 
We also show that the performance of our BS simulator is in line with the theoretical estimation of Clifford \& Clifford \cite{clifford2020faster} providing a way to define a single parameter to characterize the performance of the BS simulator in a portable way.
The developed design can be used to simulate both ideal and lossy boson sampling experiments. 
\end{abstract}

\maketitle

\section{Introduction}

The main idea behind quantum supremacy \cite{preskill2012quantum, lund2017quantum, harrow2017quantum} is to use a quantum processor to solve a mathematical problem that is intractable for the classical computers, that is the solution of the problem on classical hardware would require resources (like the execution time) scaling exponentially with the problem size \cite{aaronson_bs,Aaronson:14, bremner2016average,  boixo2018characterizing, bouland2019complexity, haferkamp2020closing, oszmaniec2020fermion,Tangpanitanon_2023}. 
Undoubtedly, it is fundamental for the researchers to ascertain that the quantum device used in the experiment indeed solves its designed task in the way it is expected.
Without such confirmation, one could not measure the strength of the quantum supremacy claims. For this reason, the development of powerful validation protocols is of high importance.

One of the proposed quantum algorithms to surpass classical computing devices is the so-called boson sampling (BS) \cite{aaronson_bs,Aaronson:14} in which bosons are drawn from a distribution of indistinguishable particles.
In case the relation $m\gg n^2$ between the number of the output modes $m$ and the number of input photons $n$ holds on, the total execution time of the simulation scales exponentially with $n$, since there are no efficient classical methods to simulate the quantum correlations originating from the peculiar nature of multi-body bosonic systems consisting of indistinguishable particles.
For this reason, BS has been subjected to numerous theoretical 
and experimental investigations.
Due to the intensive interest shown towards this topic, several flavors of BS have been formulated such as the Gaussian BS \cite{PhysRevLett.119.170501,PhysRevA.100.032326,PhysRevA.98.062322}, scattershot BS \cite{PhysRevLett.113.100502,Bentivegnae1400255,PhysRevLett.118.020502}, translating BS into time domain \cite{PhysRevLett.113.120501,PhysRevA.93.043803}, or implementing the BS with trapped ion technology \cite{PhysRevLett.112.050504,Toyoda2015}, ultra cold atoms \cite{robens2022boson} and with microwave cavities \cite{PhysRevLett.117.140505}.
From an experimental point of view, after the proof-of-principle realizations of small-scaled photonic interferometers \cite{Broome794,Spring798,Crespi2013,Tillmann2013}, researchers were experimenting to increase the number of modes on the photonic chip \cite{Spagnolo2014} and by multiplexing on-chip single-photon sources \cite{Spring:17,Faruque:18,doi:10.1116/5.0018594}. 
Currently, the largest conventional BS experiment was reported in the work of Wang et. al. \cite{PhysRevLett.123.250503} via a $60$ mode interferometer with $20$ input photons and $14$ measured photons. 
In 2020, a quantum computational advantage was claimed via sampling from a Gaussian state at the output of a $100$-mode ultralow-loss interferometer with threshold detection \cite{Zhong1460} and an average of around $45$ photons. Subsequently, this was extended to $144$ modes with (partially) programmable input states \cite{zhong2021phase}. However, these experiments were shown to be vulnerable to spoofing by drawing samples from distribution governed by classical heuristics \cite{villalonga2022efficient}.
In 2022, a further important milestone was achieved on a fully programmable photonic chip by carrying out Gaussian BS on $216$ squeezed modes using a time-multiplexed and photon-number-resolving architecture \cite{Madsen2022} that was resistant against the spoofing approach of \cite{villalonga2022efficient}.
Because of the dedicated race to experimentally demonstrate proven quantum advantage the validation protocols also increased in their importance \cite{Fefferman2023}. 
Shortly after the BS proposal researchers tried to find a way to verify the results of the samplers. Initially, the validators used statistical tests to show that the samplers draw samples from distribution being different from classical counterparts \cite{Neville_2017} such as the mean-field approach or uniform distribution. 
During recent years, we could see the emergence of more sublime BS validation approaches \cite{Seron_2022,Chabaud2021efficient}. Some of those new ideas arise from the techniques used in a different branch of computer science, like pattern recognition \cite{Agresti_2019} or computer vision \cite{Flamini_2019}. These validators require permanent computation or access to the samples drawn from a bona fide boson sampler, which means they can benefit from a high-performance permanent computation technique. 

From a mathematical point of view, the key element in the simulation of BS is a fast evaluation of the permanent function. 
The reason for that is the connection between the permanents of specific matrices and probabilities of BS outcomes. 
Thus, in BS simulation many permanent computations are required even with the most efficiently known Clifford \& Clifford algorithm \cite{10.5555/3174304.3175276,clifford2020faster}.  
The permanent of an $n\times n$ matrix $A$ is defined by the formula
\begin{equation}
    \textrm{perm}(A)=\sum\limits_{\sigma \in S_n} \prod\limits_{i=1}^n a_{i,\sigma(i)}, \label{eq:perm_def}
\end{equation}
where  $S_n$ is a set of non-repeating $n$ element permutations of $S=\{1, ..., n\}$.
The formula has a factorial time complexity of $\mathcal{O}(n!n)\le\mathcal{O}(n^{n+1})$.
Valiant et. al.\cite{VALIANT1979189} proved that the evaluation of the permanent of a matrix containing only zeros and ones belongs to complexity class \#P (sharp P)-complete, a class which is at least as hard as NP-complete with many counting problems e.g. perfect matchings, graph colorings of size $k$, satisfiable assignments to general Boolean formula, etc. 
Unlike the determinant having quite a similar definition to Eq.~(\ref{eq:perm_def}), there is not any linear algebra simplification to calculate the permanent in polynomial time.
Currently, the most efficient approaches to calculate the permanent have a computational complexity of $\mathcal{O}(n^2\cdot2^n)$ which can be further reduced to $\mathcal{O}((n\cdot2^n)$ if data recycling of intermediate results is implemented via Gray code ordering. 
The $\mathcal{O}(n^2\cdot2^n)$ variants of the so-called Ryser\cite{ryser1963combinatorial} and the Balasubramanian-Bax-Franklin-Glynn (BB/FG) \cite{Glynn2013} formulas were benchmarked on the Tiahne 2 supercomputer in Ref.~\cite{10.1093/nsr/nwy079}, implying better numerical accuracy of the BB/FG method.   

In this work, we designed a novel scalable implementation to calculate the permanent via the BB/FG formula. 
In order to account for a collision of photons on the optical modes of the interferometer we introduce an extended version of the BB/FG formula based on the concept of the n-ary Gray code ordering \cite{Kurt_adynamic} to take into account data repetition during the permanent evaluation and improve the performance of the BS simulation. 
We also implemented the designed computational model on FPGA-based data-flow engines (DFEs) further improving the computational performance of the BS simulation provided by the Piquasso package \cite{piquasso}. 
The so-called data-flow programming model \cite{10.1145/1013208.1013209}
 utilizes a fundamentally different approach to increase the computational concurrency than traditional parallel programming models on GPU and CPU hardware.
The basic concept of data-flow programming can be explained by thinking of a stream of computational data flowing through hardware. 
Each of the utilized resources performs a fixed logical transformation on the elements of the stream, transforming a single data element in each clock cycle.
By chaining up hardware resources we end up with a data-flow computing model: while one hardware element is transforming the $i$-th element in the data stream, the next hardware element in the chain is already working on the previously transformed, $i{-}1$-th element of the stream. 
By providing a long enough data stream to the hardware one can realize an efficient parallel execution, even if the computational tasks have a high degree of data dependency.

The high-level DFE development framework of Maxeler Technologies provides an efficient programming background to design data-flow applications in terms of high-level building blocks (such as a support for fix-point arithmetic operations with complex numbers, automatic pipeline scheduling, stream-holds, memory controllers, etc.) instead of the lingering work of programming low-level VHSIC Hardware Description Language (VHDL) components.
By combining a novel permanent computational approach based on the BB/FG formula with the data-flow programming model we developed a permanent computing DFE capable of supporting exact BS simulations (with and without photon losses) up to $40$ photons.
The computational complexity of the BS simulation might be significantly reduced if photon occupation multiplicities on the optical modes are taken into consideration. 
Our DFE permanent calculator is adapted to this generalization by streaming the addends in n-ary Gray code ordering directly on the FPGA chip.
With this accessory, it took $\sim 90$ seconds per sample to draw $40$ photon samples from a random $60$ mode interferometer via four concurrently operating DFEs.

The manuscript is organized as follows: in Sec. \ref{sec:BBFG} we discuss the n-ary Gray code implementation to evaluate the permanent function with the BB/FG formula, accounting for photon occupation multiplicities on the photonic modes. 
Then, in Sec. \ref{sec:DFE} we describe the DFE implementation of the permanent calculation engine developed for FPGA chips.
Finally, in Sec. \ref{sec:BS} we provide the results of our numerical experiments on the classical simulation of BS incorporating DFE permanent calculation implementations.

\section{Evaluation of the permanent function: a novel scalable approach} \label{sec:BBFG}

The two lowest complexity methods of computing the permanent involve the formulas of Ryser \cite{ryser1963combinatorial} and BB/FG \cite{Glynn2013} when using Gray code ordering of the addends.
Ryser's approach computes the permanent as 
\begin{equation}
    \textrm{perm}(A)=(-1)^n\sum\limits_{S\subseteq \{1,...,n\}} (-1)^{\lvert S\rvert} \prod\limits_{i=1}^n \sum\limits_{j\in S} a_{i,j}. \label{eq:ryser}
\end{equation}  
The formula has an exponential computational complexity of $\mathcal{O}(n^2\cdot 2^{n})$, which is significantly reduced when the subsets $S\subseteq \{1,...,n\}$ are ordered in a way that only a single element of $S$ is changed in the subsequent $S$. 
In this case, the value of the inner sum of the matrix elements can be reused, i.e. only a single matrix element should be added or subtracted from the reused sum to obtain the new value corresponding to the next $S$. 
By reducing the complexity of the inner sum from $\mathcal{O}(n)$ to $\mathcal{O}(1)$ this way, the Ryser formula can be evaluated by an overall $\mathcal{O}(n\cdot 2^{n})$ complexity.
By a later technique reported in Ref.~\cite{nijenhuis1975combinatorial} the complexity can be further reduced by a factor of $2$ in the outer sum using  the expression:
\begin{equation}
    \textrm{perm}(A) = (-1)^{n-1}\cdot2\cdot
    \sum\limits_{S\subseteq \{1,...,n-1\}} (-1)^{\lvert S\rvert} \prod\limits_{i=1}^n \left(x_i + \sum\limits_{j \in S} a_{i, j}\right),
\end{equation}   
where $x_i$ is defined by $x_i=a_{i, n}-(a_{i,1}+a_{i,2}+\dots +a_{i,n})/2$. 
Another highly efficient method to calculate the permanent is provided by the BB/FG formula: 
\begin{equation}
    \textrm{perm}(A) = \frac{1}{2^{n-1}}\sum\limits_\delta \left(\prod\limits_{k=1}^n \delta_k\right) \prod\limits_{j=1}^n \sum\limits_{i=1}^n \delta_i a_{i,j}, \label{eq:bbfg}
\end{equation}   
where $\mathbf{\delta}=(\delta_1, \delta_2, ..., \delta_n)$ with $\delta_1=1$ and $\delta_i = \pm 1$ for $1\leq i \leq n$.  
Notice, that in contrast with the traditional definition of the BB/FG formula, in the inner sum of Eq.~(\ref{eq:bbfg}) we rather calculate the column sums of the input matrix instead of row sums. 
This choice is motivated by a practical reason implied by the following reasoning: 
regarding the unitary matrix describing the scattering in the optical interferometer, the rows of the unitary are associated with the output states, while the columns are related to the input modes. 
Non-trivial photon multiplicities on the output modes (expected to occur more often than multiplicities in the input modes) are described by repeated rows in the unitary. As we will show in Sec.~\ref{sec:repeated}, by accounting for these multiplicities one can significantly reduce the complexity of the permanent evaluation. 

The computational complexity of the BB/FG formula is $\mathcal{O}(n^2\cdot 2^{n-1})$, while it can be reduced to $\mathcal{O}(n\cdot 2^{n-1})$ if Gray code ordering is applied in the evaluation of the outer sum. 
The Ryser and the BB/FG  formulas follow quite different approaches to evaluate the permanent, also resulting in dissimilar numerical properties. In the benchmark comparison of the two methods reported in \cite{10.1093/nsr/nwy079} the authors found that the BB/FG formula gives numerically more precise results than Ryser’s formula in the context of bounded bit-sized data types. 
This finding was further justified by our numerical analysis reported in Sec.\ref{sec:accuracy} comparing the numerical accuracy of the individual algorithms with the numerical results obtained by multiple-precision floating-point number arithmetics provided by the GNU MPFR library \cite{10.1145/1236463.1236468}.

\subsection{Parallel BB/FG implementation to calculate the permanent} \label{sec:scalableBBFG}

In this section we discuss the technical foundations of our CPU implementation to calculate the permanent of a square matrix, that can be further improved to account for photon multiplicities discussed in Sec.\ref{sec:repeated}.  
Our algorithm implementing a Gray code ordered BB/FG formula (\ref{eq:bbfg}) has a computational complexity of $\mathcal{O}(n\cdot 2^{n-1})$ while minimizing the overhead of processing the logic associated with the generation of auxiliary data needed for the Gray code ordering. 
Table~\ref{table:gray}. shows an example of a so-called binary reflected Gray code sequence encoded by $3$ bits.
Via Gray code ordering in each cycle of the outer sum of Eq.~(\ref{eq:bbfg}) only one element of the $\boldsymbol{\delta}$ vector is changed, making it possible to reuse the column sum (i.e. the inner sum in Eq.~(\ref{eq:bbfg})) in the next cycle and subtract/add only elements of a single row of the input matrix (see Ref.~\cite{10.1093/nsr/nwy079} for details)

Here we note some important properties of reflected Gray code counting allowing us to efficiently implement the algorithm in a parallel environment.
First of all, the Gray code corresponding to a decimal index $0\leq i< 2^{n-1}$ is $g_i=i \oplus (i >> 1)$ where $\oplus$ is a bit-wise logical XOR operation and $>>$ is a bitshift operation. 
\begin{table}
\centering
\begin{tabular}{|c|c|}
\hline
 decimal index & Gray code \\
 \hline \hline
 $0$ & $000$ \\
 \hline
  $1$ & $001$ \\
 \hline
  $2$ & $011$ \\
 \hline
  $3$ & $010$ \\
 \hline
  $4$ & $110$ \\
 \hline
  $5$ & $111$ \\
 \hline
  $6$ & $101$ \\
 \hline
  $7$ & $100$ \\
 \hline 
\end{tabular}
\caption{An example showing a $3$-bit reflected Gray code sequence. In each row a single bit is changed in the Gray code compared to the previous row.} \label{table:gray}
\end{table}
The ability to determine the Gray code corresponding to any decimal index $i$ in a constant time implies an efficient way to evaluate the permanent in parallel execution.
Simply, we divide the set of $0\leq i< 2^{n-1}$ decimal indices into smaller contiguous subsets that can be processed concurrently on the available CPU threads.
For the first element in the subset, the column sum is initialized with $n$ arithmetic operations.
Whether an $a_{i,j}$ element is subtracted or added to the column sum is determined by the elements $\delta_i$ which are mapped to the individual bits of the Gray code: 
$\delta_i$ is considered to be $+1$ ($-1$) if the corresponding Gray code bit is $0$ ($1$).
After initializing the initial column sums on each CPU thread, additional addends to the permanent are calculated via sequential iterations on the elements of the given subset of the decimal indices $\{i_{min},\dots ,i_{max}\}$. 
To this end, we need to determine the changing bit in the Gray code and its position.
In principle, the bit mask of the changed bit is given by a bit-wise comparison of the Gray code with its prior value $g_i \oplus (g_i >> 1) \oplus (g_i-1) \oplus ((g_i-1) >> 1)$.  
However, according to the generation rule of the Gray code $g_i$ from the decimal index $i$, we can use a more efficient way to determine the changed bit.
Namely, in each cycle, the position of the changed bit in the Gray code is determined by the position of the lowest bit in the decimal index which is $1$.
Also, since in each iteration only one element of the Gray code is changed we can keep track of the "parity" of the $\boldsymbol{\delta}$ vector (i.e. whether the sum of the elements $\delta_i$ is odd or even) from cycle to cycle in an efficient way without iterating over the elements of the vector.  
We provide a reference implementation of our recursive permanent algorithm in the Piquasso Boost library \cite{piquassoboost} providing high-performance C++ computing engines for the Python-based universal bosonic quantum computer simulator framework Piquasso.
We also notice that the described method can be used to scale up the computation of the permanent over multiple computing nodes (using Message Processing Interface (MPI) protocol), so in contrast with the claim of Ref.~\cite{10.1093/nsr/nwy079} the Gray coded permanent evaluation can be efficiently turned into parallel execution.

\subsection{Handling row and column multiplicities in the input matrix} \label{sec:repeated}

In the general case, multiple photons might occupy the same optical mode at the output of the interferometer.
Since the output modes are associated with the row indices of the unitary describing the scattering process of the photons, the multi-photon collision at the output modes is mathematically described by repeated rows in the unitary.
In particular, the number for which a specific row is repeated in the unitary is identical to the number of photons occupying the corresponding optical mode. 
Following the permanent evaluation strategy encoded by the BB/FG formula it becomes clear that by having row multiplicities in the input matrix some of the addends to the permanent would show up multiple times during the calculation.
Such a complexity reduction was already reported in several previous works of Ref.~\cite{Chin2018,LUNDOW2022110990,clifford2020faster} by generalizing the Ryser formula, or Ref.~\cite{Gupt2019} by turning the permanent calculation into the evaluation of the Hafnian function accounting for multiplicities \cite{doi:10.1126/sciadv.abl9236}. 
Here we argue that it is possible to make use of the addend multiplicities and reduce the overall complexity of the BB/FG permanent calculation as well. According to our best knowledge, this improvement to the BB/FG algorithm was not reported before.

For example, let's assume that the $k$-th output mode ($k>1$) is occupied by $M_k$ photons, resulting in $M_k$ identical rows in the unitary at row indices $i_k, i_k+1,\dots, i_k+M_k-1$. 
Consequently, the
\begin{equation}
    \sum\limits_{i=i_k}^{i_k+M_k-1} \delta_i a_{i,j} = \left(\sum\limits_{i=i_k}^{i_k+M_k-1} \delta_i\right) a_{i_k,j} = \left(M_k-2\Delta_k\right)\; a_{i_k,j} \label{eq:Deltak}
\end{equation}
column sum might result in $M_k+1$ different outcomes depending on the number $\Delta_k$ of $-1$ values and the number $M_k-\Delta_k$ of $+1$ values among the $\delta_i$ ($i_k\leq i \leq i_k+M_k-1$) elements of the $\boldsymbol{\delta}$ vector. 
The individual outcomes occur explicitly $\binom{M_k}{\Delta_k}$ times during the permanent evaluation.
Taking the $M_k$ multiplicities for each optical output mode labeled by $k$, in total 
\begin{equation}
    C=\prod\limits_{k=1}^{\#modes}(M_k+1)
\end{equation}
different outcomes of the inner column sum show up during the calculation, determining the overall complexity of the permanent evaluation.
According to the BB/FG formula, the first row ($k=1$) of the matrix is always taken by a coefficient $\delta_1=1$, hence we always treat the first row with a multiplicity of $1$, even if it is repeated in the matrix.
Computationally the best practice is to move one of the non-repeating rows (if exists) to the top of the input matrix.
The possible computational speedup achieved via the outlined combinatorial simplification depends on the specific use case. 
The exponential factor in the evaluation complexity is determined by the number of the involved rows of the input matrix. 
In general, the complexity of the calculations can be reduced to 
\begin{equation}
    \mathcal{O}\left( n\cdot 2^{M^*} \right), \qquad M^*=\textrm{log}_2(C)=\sum\limits_{k=1}^{\#modes}\textrm{log}_2(M_k+1),
\end{equation}
with $n=\sum M_k$ over the optical modes.
In order to account for row multiplicities within the BB/FG formula we make use of reflected n-ary Gray code counting instead of the binary Gray code counting described in Sec.~\ref{sec:scalableBBFG}:
\begin{equation}
    \text{perm}(\boldsymbol{A}, \boldsymbol{M}, \boldsymbol{N})=\frac{1}{2^{n-1}} \sum\limits_{\boldsymbol{\Delta}}  \bigg(\prod\limits_{k=1}^{\#modes} (-1)^{\Delta_k} {M_k \choose \Delta_k}\bigg) \prod\limits_{j=1}^{\#modes} \bigg(\sum\limits_{k=1}^{\#modes}
        \left(M_k-2\Delta_k\right)a_{k, j}\bigg)^{N_j}, \label{eq:bbfgrepeated}
\end{equation}
where $\boldsymbol{A}=a_{ij}$ is a square matrix describing the interferometer, $\boldsymbol{M}$ and $\boldsymbol{N}$ are the row and column multiplicities respectively such that the photon count $n=\sum M_i=\sum N_j$ and $\boldsymbol{\Delta}$ is the n-ary Gray code, required for efficient computation.
[For the physical meaning of $\Delta_k$ see Eq.~(\ref{eq:Deltak}).]
The n-ary Gray code is also known as the non-Boolean Gray code or Guan code \cite{Guan1998GENERALIZEDGC}. 
As the name implies, the individual "digits" of the n-ary Gray code are encoded by numbers from $0,1,\dots, n-1$ instead of the binary values $0$ and $1$.
While the limits of the individual Gray code digits can be different from each other, one can construct a specific n-ary Gray code counter in which the $k$-th digit counts the number of $+1$ values of the $\delta_i$ elements corresponding to the repeated rows describing the $k$-th output mode. 
Thus, according to our reasoning, the $k$-th digit of the Gray code counts from $0$ to $M_k$.
In order to construct such a reflected n-ary Gray code counter, we followed the work of Kurt et. al. \cite{Kurt_adynamic}.
In each iteration, only a single digit changes its value, enabling one to reuse the calculated column sums in the next iterations, similarly to the case of binary-reflected Gray code ordering.

Due to the reflected nature of the counter, it is possible to determine the Gray code corresponding to a specific decimal index $0\leq i<C$ in a constant time (for details see \cite{Kurt_adynamic}).
Thus, the algorithm can be executed in parallel in a similar fashion then we described it for a binary-reflected Gray code counter.
In the Piquasso Boost library, we provide our implementation of the BB/FG algorithm accounting for row multiplicities.

\section{DFE Design and Implementation to evaluate the permanent} \label{sec:DFE}

In order to further increase the computational speed of permanents, we developed a DFE implementation of the algorithm described in the previous section.
Here in this section, we discuss the technical aspects of the developed FPGA (Field Programmable Gate Arrays) based permanent calculation engines realized via a static data-flow programming model. 
Since the configuration of the programmable hardware elements on the FPGA chip (or in other words the uploading of the program to the FGPA card) is time-consuming, the implementation needs to be general enough to avoid the need to re-configure the FPGA card during the BS runtime.
In addition, by accounting for the particular needs implied by physics working behind the simulated BS architecture, we encountered further optimization requirements for the DFE implementation, such as the possibility to calculate the permanents of multiple matrices in one shot to amortize the initialization overhead even at small matrix sizes.
In order to provide a thorough description of our implementation, while sparing the reader from too many low-level technical details, we will discuss these optimization details separately in the Appendix.
Here we rather focus on the description of the basic concept to evaluate permanents on DFEs. 

Due to high resource requirements associated with floating point operations in the FPGA chip, we focused on designing a scheme where fixed point number representation was used to do arithmetic operations with gradually increasing the bitwidth of the number representation from the initial $64$ bits (sufficient to represent the elements of the input unitary) up to $128$ bits used to derive an accurate final result up to a unitary size of $40\times 40$. 
(While the floating point units on modern CPUs are highly specialized and optimized, the FPGA with the aid of generic look-up tables (LUTs) and digital signal processing (DSP) units scales better with fixed point operations. For example, double precision floating point additions require a delay of $14$ clock ticks, while $1$ tick is sufficient for fixed point addition.)
Starting with $64$ bit double precision format of the input unitary on the CPU side, we perform a conversion of floating point numbers ($f=s\cdot 2^e$) into fixed point representation encoded by $b=64$ bit variable $a$, such that $a$ is the nearest integer to $f\cdot 2^{b_f}$, with $b_f=b-2$ standing for the number of fractional bits. The remaining two bits are used to store the integer part (for the case if the matrix element is unity) and the sign of $a$ in two's complement encoding. (Since the input matrix is expected to be unitary, the magnitude of the matrix elements would never be larger than unity.)
\begin{widetext}
\begin{figure}[!h]
    \centering
    \includegraphics[width=\textwidth]{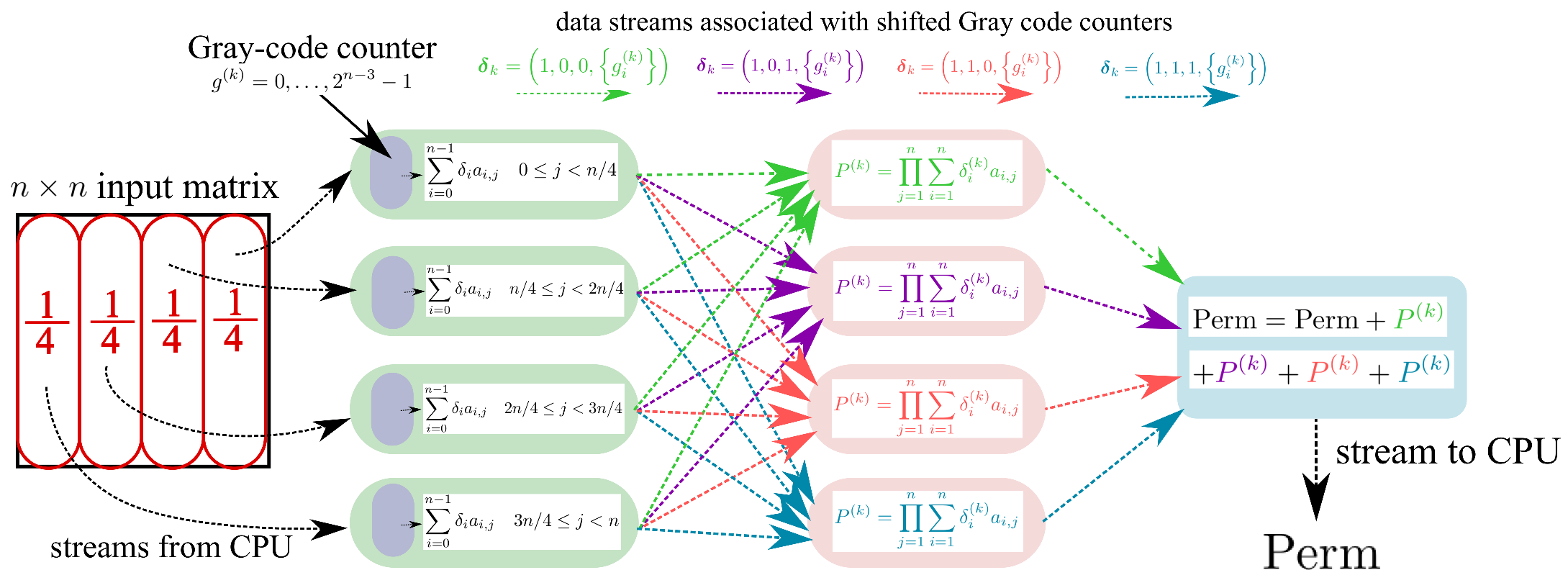}
    \caption{The structure of the computing kernels realized on the FPGA chip. The Xilinx Alveo U250 chip is organized into four super logic regions (SLRs) \cite{alveoU250}, with limited inter-SLR connectivity. 
    The kernels are organized accounting for this specific hardware setup introducing $4$ concurrent kernel blocks (mostly) bounded to the individual SLRs.
    The individual kernels are indicated with colored regions, also showing the mathematical operation they are evaluating.
    Further details of the data-flow scheme are discussed in Sec.\ref{sec:DFE}.}
    \label{fig:DFE_design}
\end{figure}
\end{widetext}

In order to ease the congestion on the FPGA chip and avoid timing issues associated with long routing paths, we split the input matrix into four blocks (see the left side of Fig.~\ref{fig:DFE_design}), and stream the column blocks into $4$ distinct "column summing" DFE kernels indicated by green colored boxes in Fig.~\ref{fig:DFE_design}).
This way the computations could be partitioned over the Xilinx Alveo U250 FPGA chip used in our numerical experiments while increasing the computational concurrency at the same time.
The purpose of these kernels is to (i) calculate the initial column sums $\sum\delta_i a_{i,j}$ for each column in the first $n$ tick of the kernel (here $n$ is the matrix size). 
Secondly, (ii) as the initial column sums are calculated, a gray code counter $g^{(k)}\in(0,2^{n-3}-1)$ is launched (in the $n$-th clock tick) to stream the last $n-3$ elements of the $\boldsymbol{\delta}^{(k)}$ vectors defined in the BB/FG permanent formula of Eq.~(\ref{eq:bbfg}).
(The corresponding $\delta_i$ elements are given by the individual bits of $g^{(k)}$.
Further details to generate the gray code counter logic via bit-wise operations are discussed in Sec.~\ref{sec:scalableBBFG}).
The first element of each $\boldsymbol{\delta}^{(k)}$ vector is $1$ by the definition, while the second and third elements become fixed by the following reasoning: 
in order to increase the computational concurrency on the DFE the gray code counter is multiplexed in order to create $4$ concurrent streams of the $\boldsymbol{\delta}^{(k)}$ vectors by fixing the second and third elements to $(0,0)$, $(0,1)$, $(1,0)$ or $(1,1)$ as indicated by the colored  $\boldsymbol{\delta}^{(k)}$ streams in Fig.~\ref{fig:DFE_design}.
Consequently in each of the $4$ "column sum" kernels $4$ concurrent $\boldsymbol{\delta}^{(k)}$ vector streams are used to calculate the $\boldsymbol{\delta}^{(k)}$ weighted column sums.
In each clock cycle $4$ new column sums are concurrently calculated for each column index $0\leq j <n$ (distributed between the $4$ column sum kernels) corresponding to the $4$ multiplexed $\boldsymbol{\delta}^{(k)}$ streams: from the value of the most recent column sum the new value is determined by adding or sub-tracking twice the matrix element taken from row $i$. The row index $i$ corresponds to the element index where a change occurred in $\boldsymbol{\delta}^{(k)}$ compared to $\boldsymbol{\delta}^{(k-1)}$ (due to the design this index is the same for all of the multiplexed $\boldsymbol{\delta}^{(k)}$ streams). 

The calculated column sum streams are gathered into groups according to the streams $\boldsymbol{\delta}^{(k)}$ to provide an input for the next layer of computing kernels. 
The red-colored kernels in Fig.~\ref{fig:DFE_design}. calculates the products of the incoming column sum streams (each of them corresponding to a specific column index $0\leq j<n$), i.e. the stream data arriving at the same clock cycle are multiplied with each other over a binary tree reduction with a depth of $\lceil \log_2 n \rceil$.

The most expensive operations in the design are the multiplications of the complex numbers associated with the column sums.  
According to the most widely used approach, the complex multiplications are broken down as $4$ fixed point multiplications and $2$ additions as $(a+bi)(c+di)=(ac-bd)+(ad+bc)i$. 
The formula suggested by Knuth in Ref.~\cite{knuth97} allows the computation of the product as $(c(a+b)-b(c+d))+(c(a+b)+a(d-c))i$ which can be implemented with $3$ multiplications and $5$ additions.
There is also the Karatsuba-style formula of $(ac-bd)+((a+b)(c+d)-ac-bd)i$ which Knuth credits to Peter Ungar from 1963.
This approach uses the same amount of operations as the prior one.
From a pipelining perspective on the FPGA chip, the first formula is more balanced as the multiplications and additions to get the real and imaginary parts can be implemented concurrently.
However, it was shown in work \cite{doi:10.1137/0613043} that the numeric stability of the Ungar formula is better, as only the imaginary part would be affected by additional inaccuracy.
Fortunately, in the context of fixed point numbers used in our design, this aspect is of less importance.
It should be also noted that the addition resulting in the final real and imaginary components would fall within the range $[-2, 2]$ requiring an extra leading bit in the intermediate computations.
In addition, the fixed point multiplications need to be further broken down to deal with efficient tiling on the hardware units of the FPGA chip, such as digital signal processing (DSP) multipliers.

In our implementation, we followed the pioneering results of Refs.~\cite{8464809,5272296} and \cite{1388200}.
According to \cite{8464809} one can use the Karatsuba multiplication formula by optimally splitting the input multiplicands into smaller bitwidth parts being the least common multiple of the width of the utilized input bit ports of the DSP units.
(Thus, the most optimal selection of the tiling factors depends on the width of the input multiplicands.)
A slightly different approach is to work out a DSP-number optimized recursive Karatsuba-like formula for the individual input bit size configurations by following the reasoning of \cite{5272296} and \cite{1388200} being applicable for rectangular-sized tilings.
(The Virtex 5 DSP units embedded inside our Xilinx Alveo U250 FPGA cards have $18\times25$ wide input ports to perform signed multiplications.)
Since our DFE implementation is designed to handle at most $40\times40$ matrices, one needs to calculate the product reduction of $40$ complex numbers.
The tree-like reduction is performed over $6$ levels, providing a balance between resource utilization and pipelining. 
On the first level $20$ products are calculated from $64$ bit wide input fixed point values (streamed from the columns sum kernels) resulting in $79$ bit results (the remaining bits are discarded).
On the second level, the $20$ results are paired up to $10$ pairs. 
The multiplications are performed again with $79$ bit-wide results.
The third level calculates the pair-wise product of $10$ numbers resulting in $93$ bit results.
The remaining $3$ levels to reduce the final $5$ numbers are done with results of $110, 158$ and $189$ bit precision.  All values are fixed points with 2 integer bits (including the sign bit) and the remaining as fractional bits.  The final values are all accumulated at $192$ (or $256$) bits of precision, $6$ of which are integer bits.
During the development, numerous technical details were addressed to end up with an efficient strategy to multiply large bitwidth complex numbers implemented in our calculations. 
We solved many issues related to the correct management of the fan-outs of intermediate computations or to the careful choice between LUTs or DSP units to perform multiplications on small bitwidth tiles in order to reach the limits of our design being still decomposable on the FPGA chip. 
The most cumbersome limiting factor was the requirement to keep up with the numerical accuracy comparable to the CPU implementations using extended precision floating point number representation (see the Appendix for the comparison of DFE implementation to the CPU ones).
To this end, we needed to increase the bitwidth of the fixed point numbers in the multiplication binary tree as much as possible by utilizing more and more hardware resources, while keeping up with the timing constraints by designing a suitable pipeline structure on the chip.

Finally, the results of the four product kernels are streamed toward the final "sum up" kernel indicated by the blue-colored region in Fig.~\ref{fig:DFE_design}.
Due to the multiplexed four gray code streams four different column product reductions are entering the kernel in each tick. 
The incoming streams are summed up and added to the partial permanent labeled by $Perm$ in Fig.~\ref{fig:DFE_design}. 
To sum up the permanent addends into a single scalar result, a single clock tick deep data-flow loop is designed (corresponding to the addition of two fixed point numbers) by removing any registers from the underlying logic.
Finally, the result encoded by a $128+128$ bit complex number is streamed back to the CPU.

Beyond the discussed layout issues, many other optimizations were implemented to increase the potential usability of the developed permanent calculator DFE in quantum computing simulations.
The most important issue in our work was to generalize the implementation for matrices of variable size.
Since the initial uploading of the DFE program onto the FPGA card takes about $\sim 1$ minute, using the engine in realistic BS simulations without this generalization is infeasible.
In Sec.~\ref{sec:general_input}, we provide the details of our solution to generalize the DFE implementation to calculate the permanent of arbitrary-sized matrices.
By preserving the computational accuracy (see Sec.~\ref{sec:accuracy}), however, the maximal size of the input matrix for which the layout could still fit onto the chip turned to be $48\times48$ at clock frequency $300$ MHz (and $330$ MHz at maximal matrix size of $40\times40$). 
For cases where the FPGA would not tick enough to get a result, i.e. the input matrix is smaller than $3\times3$, the permanent computations are performed solely on the CPU side. 
(While such cases could be computed in one tick on the FPGA with some additional logic, the communication delay with DFE would likely exceed the cost of the computation by the CPU side.)
Also, we would like to highlight at this point another important optimization of the design. While feeding the permanent calculator DFE with more matrices sequentially (i.e. one after another) in a single execution, one can retrieve the calculated permanents for all of the matrices in one DFE execution.
Since the execution of our program on the DFE takes up an overhead of about a millisecond (including I/O data transfer and other initialization overhead while the DFE program is already uploaded to the FPGA card), it is expected to provide a further advantage for the DFE, if one can calculate the permanents for more matrices in a single execution amortizing the overhead between the matrices. 
Following this ideology from Sec.~\ref{sec:batch}, we show that significant speedup can be realized in permanent evaluation even for small matrices, speeding up the simulation of BS if a large number of smaller matrices needs to be processed.

Depending on the specific needs of individual simulation tasks, it might be advantageous to introduce more concurrency into a single evaluation of permanents by splitting the calculations between multiple DFEs.
This might be especially useful when dealing with larger matrices (up to $48\times48$ in our case) during the calculations. 
In Sec.~\ref{sec:dualDFE} we provide further details on the technical background related to scaling up permanent calculation over multiple DFEs.
In order to utilize DFEs to support the simulation of photonic quantum computers (when multiple photons can collide onto the same output mode), one also needs to implement the necessary logic accounting for row multiplicities reducing the permanent calculation complexity. This optimization is discussed in Sec.~\ref{sec:repeated}.    

Finally, we can not finish this section without mentioning that in order to prevent integer overflow during the arithmetic operations on the DFE, we need to keep all of the intermediate results of calculations within certain bounds correspondingly to the limits of the used fixed point number representations. 
This can be achieved by a well-formulated normalization strategy of the input matrices, while the normalization factors can be used to re-scale the output result received from the DFE to end up with the correct result of permanent.
We provide further details on the applied normalization strategy in Sec.~\ref{sec:normalization}.

\subsection{Performance benchmark of the permanent calculator implementations} \label{sec:numerical_results}

In this section, we provide the details regarding our performance benchmark of the DFE permanent calculator implementation compared to the fastest CPU implementations available today.
The numerical experiments were conducted using Xilinx Alveo U250 FPGA cards implementing a bitstream generated from Very High Speed Integrated Circuit Hardware Description Language (VHDL) generated from Java source by Maxeler's MaxCompiler 2021.1.
Thus the heart of the developed permanent calculator DFE is a static data-flow computation model translated into VHDL by the MaxCompiler.
In the performance benchmark, we compared the DFE implementation to the Piquasso Boost library's BB/FG parallel C++ implementation and TheWalrus v0.16.2 library's Ryser double and quad precision (implementing $128$ bit floats when GNU or Intel compiler is used to build the source, and long double precision for other compilers).
The measurements were computed by averaging the execution time of $5$ computations for matrices of size smaller than $24\times 24$, otherwise, just a single execution measurement was taken.

The CPU computations were performed on a two-socket CPU system consisting of AMD EPYC 7542 32-Core processors with two logical threads on each core and two non-uniform memory access (NUMA) nodes where we bound the computations to one of the NUMA nodes. 
(Thus the CPU benchmark was measured by the performance on $64$ computing threads.)
The FPGA communication uses a high-speed PCI Express 3.0 channel over which the input matrix and series of other input parameters are uploaded, like scalar initialization parameters for each kernel, clock synchronization data, etc. 
The frequency of the DFE implementation was $320$ MHz
The initialization time of programming the DFE when uploading the compiled circuit from the generated VHDL program takes $56.2$ seconds for a single DFE while it takes $112.9$ seconds for a dual DFE mode (i.e. splitting the permanent evaluation over two DFEs).
The uploaded bitstream program is preserved on the DFE until it is reprogrammed, hence the initial uploading time is needed to be spent only at the very start of the usage.
The initialization of the TheWalrus library is around $6$ milliseconds, while the BB/FG implementation of the Piquasso Boost library exhibits the most negligible loading delay with about $19.5$ microseconds.

\begin{figure}
    \centering
    \includegraphics[width=0.6\textwidth]{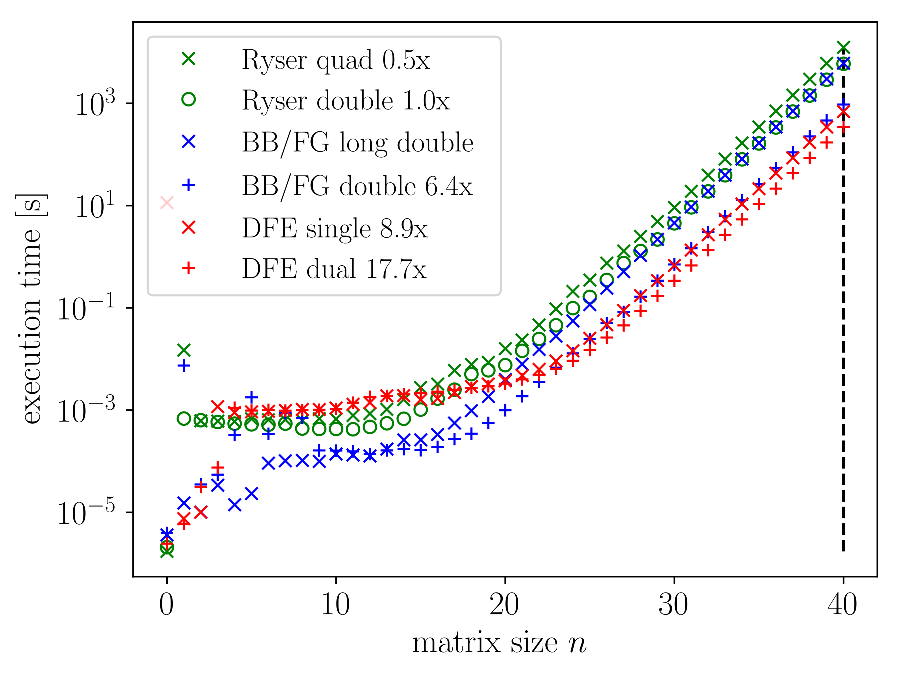}
    \caption{Performance comparison of the permanent calculator DFE to the CPU side BB/FG implementations accounting for row and column multiplicities in the input unitary.
    The input unitaries, row, and column multiplicities are constructed randomly, such that they describe $20$ (left) $30$ (middle), and $40$ photons (right) in total.
    As the photon count increases above $20$, the DFE implementation becomes systematically faster than the CPU implementations at matrix size larger than $n\approx13$. 
    The speedup factors compared to the long double precision BB/FG implementation are indicated in the legend.
    }
    \label{fig:benchmark}
\end{figure}

We compared the performance of our implementation provided in the Piquasso Boost library to the implementation of TheWalrus version 0.16.2 \cite{Gupt2019} package also having implemented parallelized C++ engines to evaluate the permanent function.
(Newer versions on TheWalrus do not contain C++ engines, nor extended precision implementations.)
The results are plotted in Fig.\ref{fig:benchmark}. showing the performance of the different libraries. 
The red-colored data points show the execution time of the single and dual DFE implementations.
(For details related to the dual mode see in Sec.~\ref{sec:dualDFE}.)
For matrices of size less than $20\times20$ the initialization overhead dominates the execution of the DFE resulting in a constant execution time.
(The permanent evaluation of the smallest matrices is done solely on the CPU side explaining the first $3$ data points in the set.)
Above the crossover region, the advantage of the DFE execution becomes evident. 
Only the double precision BB/FG implementation comes close to the performance of the DFE.
However, while the single DFE implementation is only about $1.4$ times faster than the double precision BB/FG implementation, the accuracy of the DFE implementation is significantly better, having better or identical accuracy as the long double precision BB/FG implementation up to a matrix size $40\times40$.
(For details see Sec.~\ref{sec:accuracy}.)
The long double precision CPU implementations are significantly slower than the DFE (see the indicated speedup factors in the legend of Fig.~\ref{fig:benchmark}), implying the advantage of DFE over CPU implementations when high numerical accuracy is required.
\begin{figure}
    \centering
    \includegraphics[width=0.95\textwidth]{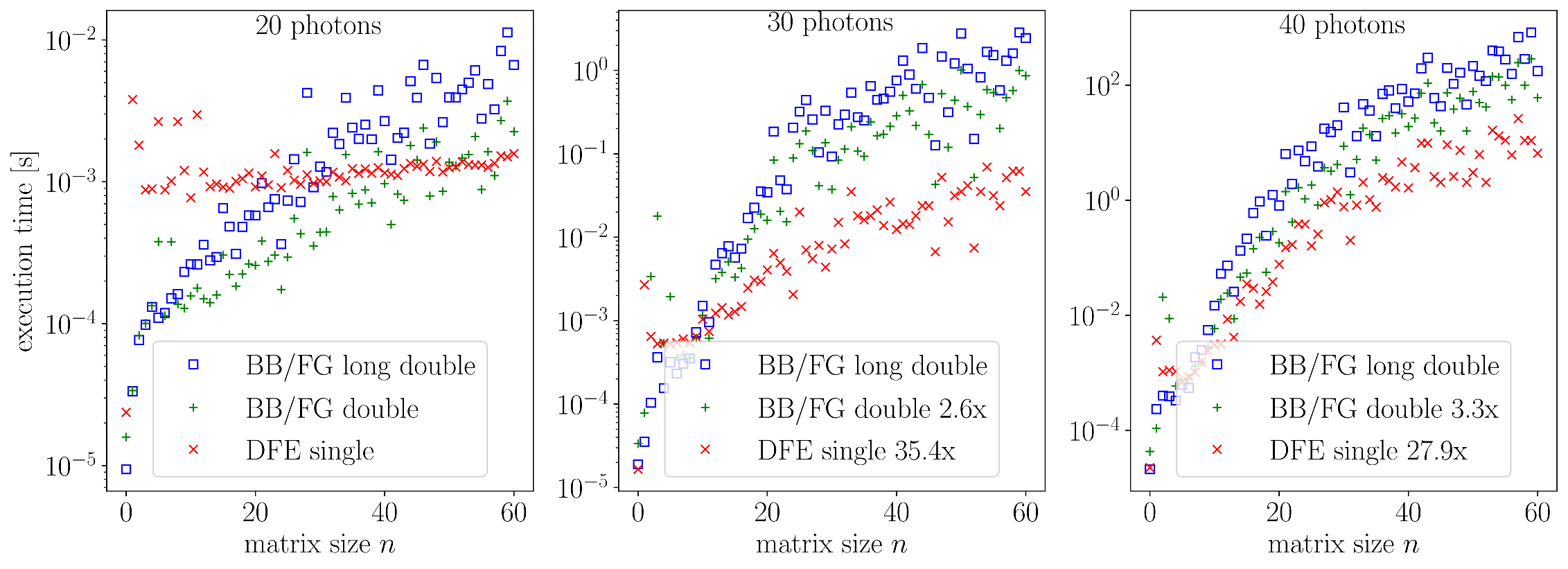}
    \caption{Performance comparison of the permanent calculator DFE to the CPU side BB/FG implementations accounting for row and column multiplicities in the input unitary.
    The $n\times n$ input unitaries, row, and column multiplicities are constructed randomly, such that they describe $20$ (left) and $30$ photons (right) in total.
    As the photon count increases above $20$, the DFE implementation becomes systematically faster than the CPU implementations at matrix size larger than $n\approx13$. 
    The speedup factors compared to the long double precision BB/FG implementation are indicated in the legend.
    }
    \label{fig:benchmark_repeated}
\end{figure}
The total run-time to evaluate the permanent of an $n\times n$ unitary on a single ($k=2$) and dual ($k=3$) DFE can be calculated as $t=t_0+\frac{n-1+2^{n-1-k}}{f}$ where $t_0$ is a small PCI Express communication delay plus the pipeline delay of the kernels (approximately half a millisecond), and $f$ is the frequency in Hertz.
The $n-1$ clock cycles are spent to initialize the column sum, and in the remaining $2^{n-1-k}$ ticks the permanent in evaluated. 
Our design is compiled for $330$ MHz frequency so the execution time for a $40\times40$ input matrix on the dual DFE build is in total $207$ seconds. (equivalent to $337$ GOPS ($10^9$ operations per seconds)
For such "long-run" execution, the $t_0$ initialization overhead is negligible.
For comparison, Ref.~\cite{10.1093/nsr/nwy079} reported the required time to compute the permanent of a $40\times40$ matrix on $98304$ CPU cores of the Tianhe 2 supercomputer in $24$ seconds using Ryser's formula in double precision.
Consequently, one CPU server with two DFE engines calculates a $40\times40$ permanent only $8.6\times$ slower than $4096$ CPU nodes (each node containing of $24$ cores) in the benchmark of Ref.~\cite{10.1093/nsr/nwy079}.

From a practical usability point of view, we also examine the performance of the permanent calculation implementations when photon multiplicities on the photonic modes induce repeated rows and columns in the unitary describing the photonic scattering process.
(See details in Sec.\ref{sec:repeated}.)
In this case, the average photon multiplicity on the photonic modes is controlled by the unitary size $n$ and the total photon number.
In Fig.~\ref{fig:benchmark_repeated}, we compare the performance of the DFE engine to the CPU implementations of the Piquasso Boost library for photon counts $20$ and $30$.
The figure shows the execution time measured at permanent evaluation for random unitaries constructed for randomly distributed photons over the photonic modes.
From our numerical results, we see that at a photon count larger than $20$ the DFE turns out to be more efficient than the CPU implementations in the evaluation of permanents. 
While keeping up to the numerical accuracy of the long double precision BB/FG implementation, the DFE is about $3.6$ times faster than the double precision BB/FG implementation at photon count $30$ and about $7.4$ times faster at photon count $40$.
(The speedup compared to long double precision implementations at the same photon counts are $35.4\times$ and $27.9\times$, respectively.)
By using dual DFE implementation the speedup is further doubled.
At lower photon count, however, the PCIe delay time of the DFE dominates the execution.
(See the left side of Fig.~\ref{fig:benchmark_repeated}.)
In the repeated row DFE variant of the permanent calculator, the columns sum initialization achieves the same performance by staggering computations the loop tiling/staggering technique.  This requires the CPU to pre-compute and upload some additional data which scales on the order of the loop length.  The loop length in practice is small or 9 cycles.  The cycle delay is based upon multiplication in computing the binomial updates.  To keep the parallelism of 4 or 8, we also force the first 3 or 4 row multiplicities to be 1 (the original formula already requires one such reduction and 2 or 3 extra for parallelism), which can slightly increase the operation count.  However, by choosing the smallest multiplicities for row expansion, the consequence of maintaining power-of-2 parallelism is largely mitigated. 
 The DFE implementation accounting for row/column multiplicities is also compiled for $330$ MHz frequency.

Finally, in Sec.~\ref{sec:batch}, we show that by streaming multiple input matrices to the DFE in a single execution, one needs to pay the PCIe delay time only once and divide it between the permanent calculation instances, thus lowering the crossover in the DFE-CPU performance down to a matrix size of $n\approx13$.

\subsection{Numerical accuracy of permanent calculator engines} \label{sec:accuracy}

As pointed out earlier by Ref.~\cite{10.1093/nsr/nwy079} the numerical accuracy becomes a notable issue with increasing the size of the input matrix. 
The final numerical precision of an implementation depends on the interplay of various factors. 
The number representation used in the individual arithmetic operations and the conversion between them, or the computational design of mathematical operations have both great effect on the accuracy of the final result.
For example, Ref.~\cite{10.1093/nsr/nwy079} found that the BB/FG formula shows bigger numerical fidelity than the Ryser formula in the experiment of reproducing the analytical result evaluated for the identity matrix  
\begin{figure}
     \centering
     \includegraphics[width=0.6\textwidth]{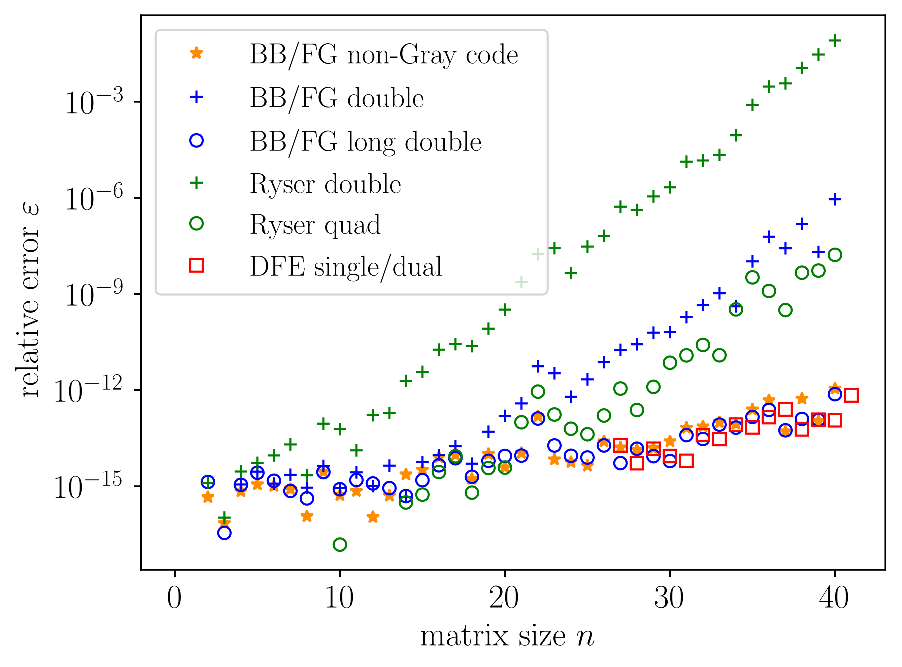}
     \caption{The numerical accuracy of different implementations to calculate the permanent function. The relative error defined by Eq.~(\ref{eq:relative_error}) was calculated for the permanent calculator engines implementing the Ryser formula implemented in TheWalrus package, the recursive BB/FG formula implemented in the Piquasso Boost library and the BB/FG formula implemented on the DFE. (For details of the DFE implementation see Sec.\ref{sec:DFE})
     Among the CPU implementations, the highest accuracy was achieved by the recursive BB/FG implementation utilizing extended precision floating point number representation.
     The implementations designed for the DFE achieve almost identical precision, we experienced deviation from the ideal result only at larger matrices.
     However, even in this regime, the relative error is smaller than $\varepsilon\approx 10^{-8}$.
    The CPU benchmark was done on an \emph{AMD EPYC 7542 32-Core Processor} platform, for DFE calculations Xilinx Alveo U250 FPGA cards were used.}
     \label{fig:accuracy}
 \end{figure}
According to our reasoning, the different numerical properties of the two approaches can be explained by the difference in the number of addends in the inner sums of Eqs.~(\ref{eq:ryser}) and (\ref{eq:bbfg}). 
While in Eq.~(\ref{eq:bbfg}) the sum involves always the same number of matrix elements before calculating the products of the column sums, in Eq.~(\ref{eq:ryser}) the number of the summed matrix elements varies according to the actual partitioning $S$ resulting in a possible wider range of magnitude in the calculated products before summing them up.
In order to increase the accuracy of the calculated permanent in the Piquasso Boost library, we evaluate the outer sum of the BB/FG formula by classifying the individual addends according to their magnitude, split them into "pockets" and always calculate the sum of those partial results that are close to each other in magnitude. 
(We also applied this strategy in our previous work \cite{kaposi2022polynomial}.)

Now we examine the numerical accuracy of the individual calculation methods.  We implemented the designed scalable BB/FG algorithm with several levels of floating point number representations provided by the Piquasso Boost library and compared the result to the Ryser formula implemented in TheWalrus package (version 0.16.2). 
Among them, the least accurate variant turned out to be the Ryser formula implemented by double precision (i.e. 64 bit) floating point arithmetic operations, followed by the double-precision BB/FG formula evaluated by the Gray code strategy. 
As default, the Piquasso Boost library implements extended precision floating point arithmetic operations to calculate the permanent providing high numerical accuracy for photonic quantum computer simulations up to photon numbers still accessible on traditional HPC nodes. 

To establish a proper ground truth, we also incorporated the GNU Multiple Precision (GMP) arithmetic library's \cite{10.1145/1236463.1236468} extension of Multiple Precision Floating-Point Reliability (MPFR) to compute the permanent with "infinite precision" using the designed recursive BB/FG algorithm. 
(The correctness of the "infinite precision" implementation was tested on special input cases. 
namely, when evaluating the BB/FG formula on $m\times n$ rectangular shaped matrices with $m\geq n+2$, the final result should be inevitably $0$ due to the exact cancellation of the addends. Such a computation, where approximation with normal floating or fixed point numbers would never return a proper $0$ result, provides very convincing evidence for the correctness of the implementation.)
Figure \ref{fig:accuracy} shows the relative error 
\begin{equation}
\varepsilon = \frac{\textrm{abs}\big(\textrm{perm}_{INF}(\mathbf{A})-\textrm{perm}(\mathbf{A})\big)}{\textrm{abs}(\textrm{perm}_{INF}(\mathbf{A}))} \label{eq:relative_error}
\end{equation}
of several benchmarked permanent calculation implementations with $\textrm{perm}_{INF}(\dots)$ standing for the infinite precision implementation.
Our numerical experiments carried out on random unitary matrices justify our expectations. 
(This choice is justified as our primary goal is to use the developed permanent calculation engines to support photonic quantum computer simulations, where the physical system is described by unitary matrices.)
The $64$ bit floating point representation is significantly less accurate than the extended precision counterparts.
Secondly, our results revealed that the Ryser formula is less accurate than the BB/FG variant.
(The accuracy of the double precision Gray coded BB/FG implementation is close to the Ryser formula evaluated with extended precision.)
A reduction in accuracy associated with the extended precision Ryser method first appears at a matrix size of $n\approx20$ and the difference increases with the matrix size. (See the green circles in Fig.~\ref{fig:accuracy}.)
Though we leave the discussion of the implementation details of the DFE permanent calculator engine for Sec.~\ref{sec:DFE}, here we just notice that the numerical accuracy of the DFE implementation stays close to the extended precision CPU implementation of the BB/FG formula, we experienced a deviance only at matrix size $n\geq36$. 
However, even at such matrix size the accuracy of the DFE implementation still remains better than the extended precision Ryser formula.

According to our reasoning, the main source of the numerical error in the CPU implementations can be explained by the fact that in Gray code variants of the Ryser and BB/FG formula, some computational data are reused from cycle to cycle, in each turn modifying their value by addition/subtraction. 
Consequently, some addends to the permanent are derived via exponentially many arithmetic operations.
This results in an accumulation of numerical error compared to the naive $\mathcal{O}(n^2\cdot2^n)$ implementations, where each addend added to the permanent is a result of only $n^2$ arithmetic operations.
This reasoning leads us to the conclusion that the $\mathcal{O}(n^2\cdot2^n)$ implementations are expected to be more accurate than the Gray-coded $\mathcal{O}(n\cdot2^n)$ variants.
We justified our expectation by evaluating the numerical accuracy of the BB/FG formula without the Gray code strategy. 
The associated orange-colored data points in Fig.~\ref{fig:accuracy} revealed, that the double precision, non-Gray-coded BB/FG formula indeed gives as good accuracy as the Gray coded variant BB/FG formula evaluated with extended precision.
The Gray-coded implementations, in turn, perform so much faster that executing them in extended precision (to obtain equivalent numerical precision) is still favorable.

\section{Classical simulation of Boson Sampling including photon losses} \label{sec:BS}

\begin{figure}
    \centering
	\includegraphics[width=0.9\linewidth]{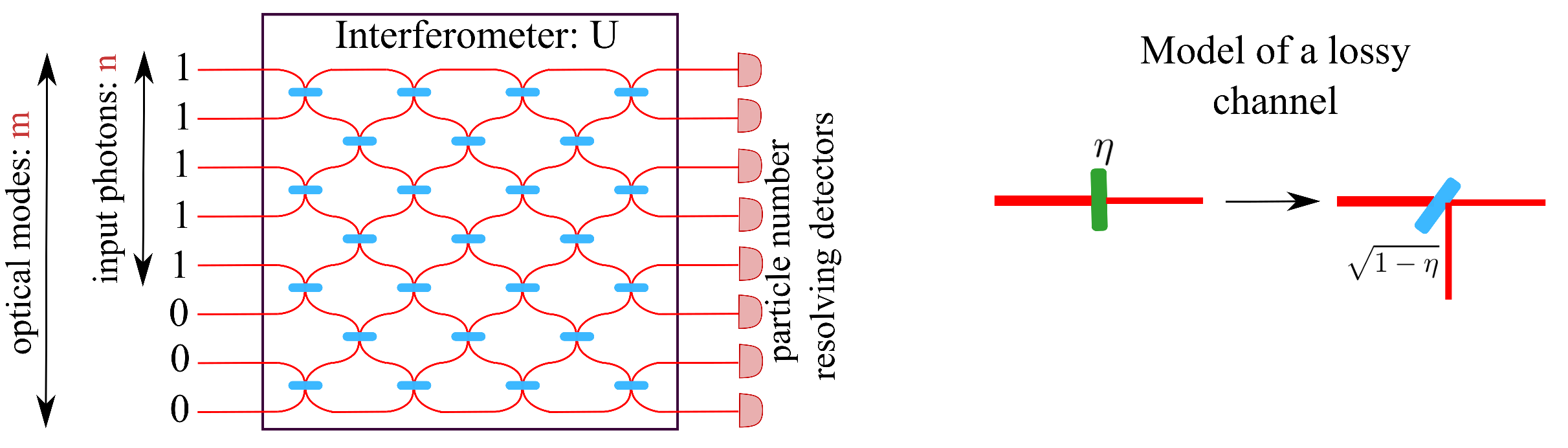}
	\caption{\textit{Left:} A schematic of a standard BS experiment. The blue elements denote beam splitters inside the interferometer. A net of interconnected beam splitters is the usual physical realization of the interferometer. \textit{Right:} A schematic representation of the loss model we use in this work. On the left side of the diagrammatic equation, we see a lossy channel on a single mode, with particle transmission probability $\eta$. On the right side of the equation, one can see the applied beam splitter that mimics such a channel. The transmitted beam labeled by $\sqrt{1-\eta}$ is not measured.}
	\label{fig:bs}
\end{figure}

The classic experimental setup of BS is shown in Fig.\ref{fig:bs}. Given an $m$-mode $n$-particle Fock state $\ket{\vec{S}}=\ket{1...10...0}$, and the interferometer, described by a $m \times m$ matrix $U$, one performs the passive (particles-number preserving) linear-optical evolution of $\ket{\vec{S}}$ with the interferometer, then we measure the outcome using the particle-number resolving detectors.

In subsequent formulas, we use $s_i$ and $t_i$ to label the components of the occupation vector describing the input state $\vec{S}$ and the output state $\vec{T}$ as $\vec{S}=(s_1=1, s_2=1,  ..., s_n=1, s_{n+1} = 0, ..., s_m=0)$ and $\vec{T}=(t_1, t_2,\dots,t_m)$.
In the particle-conserving case, $\sum_{i=1}^m s_i = \sum_{i=1}^{m} t_i$. 
The output probability $p_U$ associated with a given process $\vec{S} \to \vec{T}$ is then given by \eqref{eq:bs_p}
\begin{equation}
	\label{eq:bs_p}
	p_U(\vec{S} \to \vec{T}) = \frac{|\textrm{perm}(U_{ST})|^2}{\prod^m_{i=1} s_i! t_i!},
\end{equation}
where $U_{ST}$ is an $n \times n$ matrix constructed from the unitary $U$ describing the interferometer as follows. First, we construct an $m \times n$ matrix $U_S$, by taking $i$-th column of $U$ $s_i$ times and then we take the $j$-th row of $U_{S}$ $t_j$ times.
The hardness of BS is a consequence of the fact that the probability amplitude in Eq. \eqref{eq:bs_p} is proportional to matrix permanent. The situation is especially clear in the regime $m\gg n$, where for typical Haar random $U$ the probability distribution given in Eq.~\eqref{eq:bs_p} is concentrated on collision-free sub-spaces spanned by states $\ket{\vec{T}}$ with $t_i\geq 1$. This is the regime in which arguments for hardness put forward in the original BS paper \cite{aaronson_bs} hold (see also \cite{NoiseFrontier2021} for a more refined analysis of hardness of this problem). The hardness of BS in the regime when the number of photons is comparable to the number of modes requires a separate analysis and is a subject of an upcoming work \cite{BosonSamplingMN}.

From an experimental point of view, it should be noted that in real-world devices realizing BS photon losses occur on a regular basis. Thus, 
\begin{equation}
	\label{eq:particles_loss}
	n = \sum_{i=1}^m s_i \geq \sum_{i=1}^{m} t_i
\end{equation}
where we previously defined all of the symbols. 
Lossy BS introduces many new challenges in simulations. 
In \cite{garcia2019simulating} and \cite{Brod2020classicalsimulation} the authors discuss loss handling and a classical simulation algorithm for lossy BS simulation.
In general, interferometers are a net of interconnected beam splitters. In the lossy scenario, the lossy beam splitter transfers the particle into an inaccessible (unmeasurable) mode. One can easily see the practicality of this approach as its implementation doesn't need any new tools besides beam splitters. Graphically, one usually presents it as in the right panel of Fig.~\ref{fig:bs}.
The authors of \cite{Brod2020classicalsimulation} also noticed that the uniform part of the losses (i.e. loss that applies uniformly to all of the modes) can be extracted from the interferometer and applied as a pure-loss channel at the beginning of the simulation leading to a new instance of the matrix $U$ describing the interferometer. One may still incorporate further, non-uniform losses into the model, while the simulation will be computationally less demanding due to the lower number of particles at its input. It's worth pointing out that the matrix describing a lossy interferometer is no longer unitary as its eigenvalues can be lower than unity.

The most popular algorithm of BS simulation is the Clifford \& Clifford algorithm \cite{10.5555/3174304.3175276,clifford2020faster}. Although it was designed for the exact simulation of BS, the algorithm can be adopted for lossy BS simulations as well by virtually doubling the number of the optical modes according to the reasoning of Ref. \cite{garcia2019simulating}, while taking the sample only from the first half of the outputs.

We implemented the Clifford \& Clifford algorithm in the Piquasso Boost high-performance C++ library linked against traditional CPU and DFE permanent calculation engines. 
The BS experiment can be designed by the high-level programming interface of the Piquasso framework in Python language.
The BS simulation can be scaled up over multiple servers via MPI communication protocol. 
In our numerical experiments, we used \emph{AMD EPYC 7542 32-Core Processor} servers, each server containing $2$ CPU nodes and two Xilinx Alveo U250 FPGA cards. 
We executed the BS simulation on $4$ MPI processes, each process utilizing one CPU node with one FPGA card.
\begin{figure}
    \centering
	\includegraphics[width=0.75\linewidth]{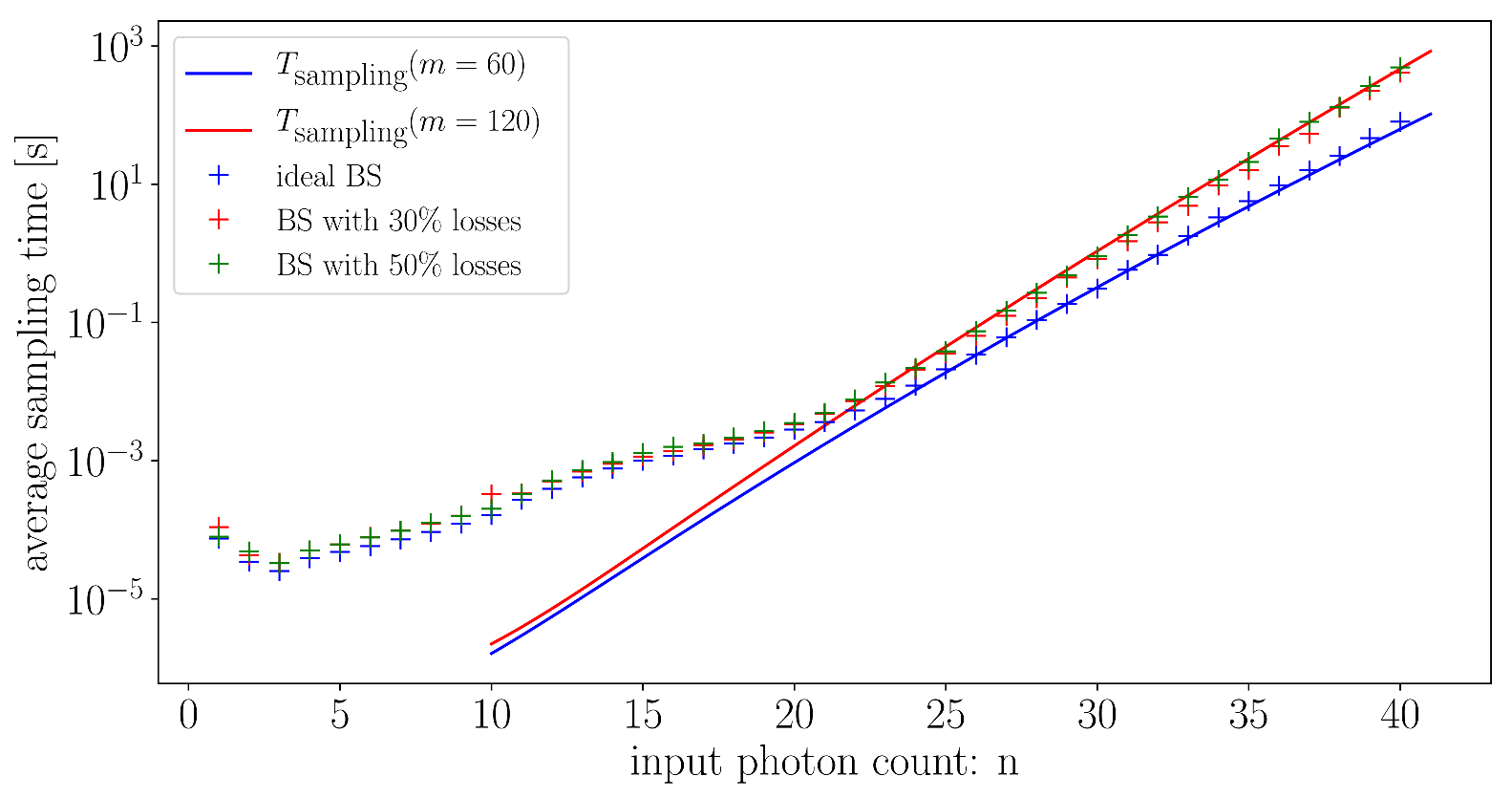}
	\caption{Performance of the developed BS simulation design executed on two \emph{AMD EPYC 7542 32-Core Processor} servers, each server containing of $2$ CPU nodes and two Xilinx Alveo U250 FPGA cards. The averaged sampling time $T_{\textrm{sampling}}$ of a $60$-mode BS simulations calculated over $1000$, $300$, $100$ samples for input photon counts less or equal than $34$, $38$, $40$, respectively.}
	\label{fig:bs_benchmark}
\end{figure}
Figure \ref{fig:bs_benchmark} shows the results of our BS simulation performance measurement carried out on host servers equipped with the DFE permanent evaluation engines.
Following the reasoning in Clifford \& Clifford \cite{clifford2020faster}, the theoretical prediction of the average sampling time (i.e. averaged over many samples) in the BS simulation (proportional to the average-case sampling complexity) can be given by the
\begin{equation}
    T_{\textrm{sampling}} = T_0 \left[ \frac{n(m+n)}{m} \left(\begin{array}{c}
    m+n \\ n+1 
    \end{array}\right)^{-1}   \left(\begin{array}{c}
    2m+n \\ n+1 
    \end{array}\right)  + n^2m \right] \label{eq:performance}
\end{equation}
expression in the $n,m\rightarrow\infty$ limit. 
The expression contains a single parameter $T_0$ providing a portable measure to capture the performance of a BS simulator.  
We fitted Eq.\~(\ref{eq:performance}) to three data sets obtained from the averaged sampling time of a $60$-mode interferometer. 
One data set corresponds to an ideal BS without any photon loss in the simulation. 
The second and third data sets were obtained by simulations assuming $30\%$ and $50\%$ losses.
(The $60$-mode interferometer and $30\%$ loss correspond to the experimental setup of Ref.~\cite{PhysRevLett.123.250503}.)
In the case of lossy BS, the simulation implies a doubling of the photonic modes, resulting in increased sampling time reported also in Fig.~\ref{fig:bs_benchmark}.
We found that Eq.~(\ref{eq:performance}) describes the complexity of BS simulations very well at larger input photon counts. 
For the developed BS simulator design we obtained $T_0 = 7.5\times 10^{-11} \;\textrm{sec}$ from the fitting.
At lower input photon count, the measured sampling performance is different from the prediction of Eq~(\ref{eq:performance}). 
At the corresponding $10^{-3}$--$10^{-6}$ s timescale the CPU-DFE performance crossover, the CPU parallelization overhead, and the initialization cost of the computational model (such as the Python-C++ binding, library loading, etc.)
dominates the execution time of the BS simulator.

Finally, we notice that the obtained $T_0$ fitting parameter is expected to be inversely proportional to the number of the incorporated DFEs, scaling ideally up to a large number of accelerators. Since the CPU servers hosting the DFEs are collecting the samples on their own, the gathering of the samples over the host servers will pose a single-time communication overhead that can be neglected compared to the overall sampling time. 
We have successfully demonstrated this design using two CPU servers hosting $4$ FPGA cards in total. 

\section{Discussion}

In this work, we described a novel scalable approach to evaluate the permanent function based on the BB/FG formula.
Utilizing the mathematical properties of reflected Gray code ordering one may concurrently evaluate partitions of the outer sum, paving the way for parallel computation of the permanent.
We further generalized the algorithm for cases when columns or row multiplicities occur in the input matrix (corresponding to multiple occupations of photonic modes) and the complexity of the permanent calculation might be reduced.
We achieved this by the utilization of generalized n-ary Gray code ordering of the outer sum in the BB/FG permanent formula, with digits varying between zero and the occupation number of the individual optical modes. 
This generalization makes the BB/FG formula applicable for high-performance BS simulation utilizing significant reduction in the computational complexity as it was previously demonstrated using the Ryser formula as well \cite{Chin2018,LUNDOW2022110990,clifford2020faster}.
The main advantage of the BB/FG formula opposed to the Ryser variant lies in the numerical accuracy of the calculated permanent value. 
Our numerical experiments using the MPFR multi-precision library showed that Ryser's formula loses against the BB/FG method by several orders of magnitude in accuracy in both the double and extended precision calculations.

We also implemented the BB/FG method on FPGA-based data-flow engines.
Our implementations are capable of handling matrices of arbitrary size (up to $40$ columns) without the overhead of reprogramming the FPGA chips and accounting for row multiplicities in the input matrix via the N-ary Gray code ordering ported to the FPGA chips.
The throughput of the DFEs was further increased by organizing the input matrices into batches and executing multiple permanent calculations on the FPGA chips in a single execution.
Finally, the fix-point number representation implemented on the FPGA chips provides competitive accuracy to the extended precision BB/FG CPU implementation in the evaluation of the permanent of unitary matrices. 
The accuracy of the DFE implementation -- equivalent to extended precision -- holds up to matrices of size $n=40$.

These features turned out to be essential to achieve an efficient BS simulator design supported by high-performance DFEs. 
We integrated our permanent evaluation DFEs into the computation flow of both the ideal and lossy BS simulation.
Since the simulation of lossy BS involves twice as many photonic modes as the ideal variant of the same BS design, the simulation of lossy BS takes more time in general. 
On average, our setup of $4$ Alveo U250 FPGA cards made it possible to take a single sample from a $60$-mode interferometer with $40$ input photons in $\sim 80$ seconds without photon losses.
Introducing photon losses into the design, our numerical experiments could draw a single sample in $\sim 360$ seconds.
The theoretical description of Ref.~\cite{clifford2020faster} fits the measured performance data very well by fitting a single parameter (labeled by $T_0$ in Eq.~(\ref{eq:performance}) to all data points.
The fitting parameter provides a portable measure to compare different BS simulator implementations. 
However, we did not find any competitive work in the literature on BS simulation providing similar performance measurements to ours.
In turn, we can compare the performance of our simulator to a real BS experiment involving $60$ photonic modes, $20$ input photons, and $14$ measured photons (i.e. loss of $30\%$ on average).
We could simulate the described design in $\sim 0.8$ milliseconds per sample, while in Ref.\cite{PhysRevLett.123.250503} authors detected $150$ valid samples of $14$-photon coincidence measurements in $26$ hours.

Finally, we notice that the BS simulation capabilities described in this work can be further improved by utilizing the concept of approximate BS described in Ref.~\cite{Brod2020classicalsimulation}, in which part of the optical modes are treated with MF approximation. 
In this approach, the number of the approximated modes is a hyper-parameter in the algorithm controlling both the speed and the fidelity of the BS simulation.
We leave the study of approximate BS simulation with DFEs for future work.

\section{Acknowledgements}

This research was supported by the Ministry of Culture and Innovation and the National Research, Development and Innovation Office within the Quantum Information National Laboratory of Hungary (Grant No. 2022-2.1.1-NL-2022-00004), by the ÚNKP-22-5 New National Excellence Program of the Ministry for Culture and Innovation from the source of the National Research, Development and Innovation Fund, and by the Hungarian Scientific Research Fund (OTKA) Grants No. K134437 and FK135220.
RP. acknowledges support from the Hungarian Academy of Sciences through the Bolyai J\'anos Stipendium (BO/00571/22/11) as well.
We acknowledge the computational resources provided by the Wigner Scientific Computational Laboratory (WSCLAB) (the former Wigner GPU Laboratory). TR and MO acknowledge financial support by the Foundation for Polish Science through TEAM-NET project (contract no. POIR.04.04.00-00-17C1/18-00).

\bibliographystyle{unsrt}
\bibliography{references}

\onecolumn\newpage
\appendix

\section{Technical details of the DFE implementation to efficiently evaluate the permanent}

\subsection{Generalization to matrices of variable size} \label{sec:general_input}

A fundamental concept of a DFE is that the static data-flow circuit implemented on the chip can not be modified during the execution.
This limitation is a crucial issue when it comes to practical QC simulation use cases of processing matrices of variable size.
In these scenarios, the $\sim1$ minute-long reconfiguration of the FPGA chip is not a viable solution either.
Fortunately, the BB/FG formula implemented in the DFE has a favorable mathematical property enabling us to generalize our DFE implementation to accept matrices of arbitrary size (up to $48\times48$).

We explain our solution by an example of a $3\times3$ input matrix, for which we are about to calculate the permanent.
We embed this square-shaped matrix into a rectangular matrix of size $3\times48$, having $3$ rows and $48$ columns:
\begin{equation}
B = \begin{pmatrix}
a_{11} & a_{11} & a_{11} & 1 & \dots & 1 & 1 \\
a_{21} & a_{22} & a_{23} & 0 & \dots & 0 & 0 \\
a_{31} & a_{32} & a_{33} & 0 & \dots & 0 & 0 \\
\end{pmatrix}
\end{equation}
The matrix $B$ contains the $a_{ij}$ elements of the original matrix, while the remaining elements of the first row are set to unity.
The remaining elements of the rectangular matrix are set to zero.
The BB/FG formula (\ref{eq:bbfg}) can be generalized for rectangular-shaped input matrices, just the upper limit of the product reduction of column sums is needed to be adjusted. 
Due to the chosen structure of zeros and ones in matrix $B$, the column sums calculated for the corresponding columns would be equal to unity for all of the vectors $\boldsymbol{\delta}$.
These column sums would not contribute to the product reduction in Eq.~(\ref{eq:bbfg}) and the final result of the formula would be mathematically identical to the permanent of the initial $3\times3$ sub-matrix as if there were no additional columns added to the matrix.
By streaming the padded matrix to the DFE, all of the data flowing through the pipelines would be well-defined resulting in the correct result for the permanent.
Thus, we can use the DFE implementation to calculate the permanent of any unitary of size less than or equal to $48\times48$ without increasing its computational time.
The upper limit $48$ for the input matrix size comes from the amount of the available hardware resources on the FPGA chip used in the project. (For details see Sec.~\ref{sec:DFE}.)  

\subsection{Dual DFE implementation} \label{sec:dualDFE}

In order to increase the concurrency in permanent evaluation we show that it is possible to split the 
calculations between multiple DFEs.
To this end, we follow the same methodology already used to increase the number of computing kernels on a single DFE.
In Sec.~\ref{sec:DFE}, we described our Gray code based strategy to create concurrent streams of vectors $\boldsymbol{\delta}$ by fixing the three most significant bits of the individual streams determining the elements of $\boldsymbol{\delta}$.
(See the colored data streams in Fig.~\ref{fig:DFE_design}.)
In order to expand the calculations over two DFEs, we increased the number of concurrent $\boldsymbol{\delta}$ streams from $4$ to $8$ and fixed the four leading bits in the corresponding streams.
The values retrieved from the two DFEs are then summed up on the CPU side giving the final result for the permanent.
Since on each DFE the amount of the processed $\boldsymbol{\delta}$ vectors becomes reduced by a factor of $2$ compared to a single DFE implementation, the execution time would be also reduced by the same factor.
To be more precise, both of the DFEs need to initialize the column sums in the first few clock cycles, implying an overhead for the parallel scaling of the calculations.
However, the overhead associated with column sum initialization is negligible in most cases and the scaling of the calculations over multiple DFEs is close to ideal.
The approach to split computations between DFEs can be increased to more DFEs as well.

\subsection{Accounting for row multiplicities in the DFE implementation} \label{sec:repeatedDFE}

Following the ideology described in previous sections, in order to decrease the computational time of permanents with repeated rows and columns we need to implement the additional logic associated with Eq.~(\ref{eq:bbfgrepeated}) on the FPGA chip as well. 
However, due to the peculiar properties of the FPGA hardware resources, several additional considerations and practices needed to be incorporated into the strategy of Sec.~\ref{sec:repeated}.
Namely, the calculation of the binomial coefficients
\begin{equation}
    b_k = {M_k \choose \Delta_k} 
\end{equation}
in Eq.~(\ref{eq:bbfgrepeated}) is the main bottleneck of a working DFE implementation.
The issue of the variable number of terms to calculate the individual $b_k$-s within static data-flow context, and the increment of pipelining depth associated with the multiplications and divisions are just two examples indicating the challenge of generalizing the DFE implementation to repeated rows/columns permanents. 
In order to decrease the resource requirements associated with the calculation of the binomial coefficients, we make use of the data-recycling approach of the Gray-coded implementation and calculate the binomial coefficient from its value in the previous iteration. 
Since the N-ary Gray-code counter changes its value only at a single place by $\pm1$ in each step, the binomial coefficient corresponding to the changed element $k$, can be calculated as
\begin{equation}
    b_k^{(\textrm{next})} = b_k\frac{M_k-\Delta_k}{\Delta_k + 1}, \quad\textrm{or}\quad b_k^{(\textrm{next})} = b_k\frac{\Delta_k}{M_k-\Delta_k}, \label{eq:simplified_binomial}
\end{equation}
when the element of the Gray-code counter changes by $-1$ or $+1$, respectively.
Since all the other binomial coefficients remain unchanged, the overall multiplicity factor (i.e. the product of the binomial coefficients) can be calculated in a fashion of Eq.(\ref{eq:simplified_binomial}).
Thus, the calculation of the multiplicity factors is simplified to a single multiplication and division and some extra logic to predict the element of the N-ary Gray code to be changed (and the value of the change).
This additional logic is implemented only in a computing kernel closest to the PCIe port of the FPGA chip to avoid long-range routing of extra data needed for the calculations. 
Finally, the multiplicity factor is distributed to all of the product kernels to be utilized in the evaluation of the permanent.

\begin{figure}
    \centering
    \includegraphics[width=0.55\textwidth]{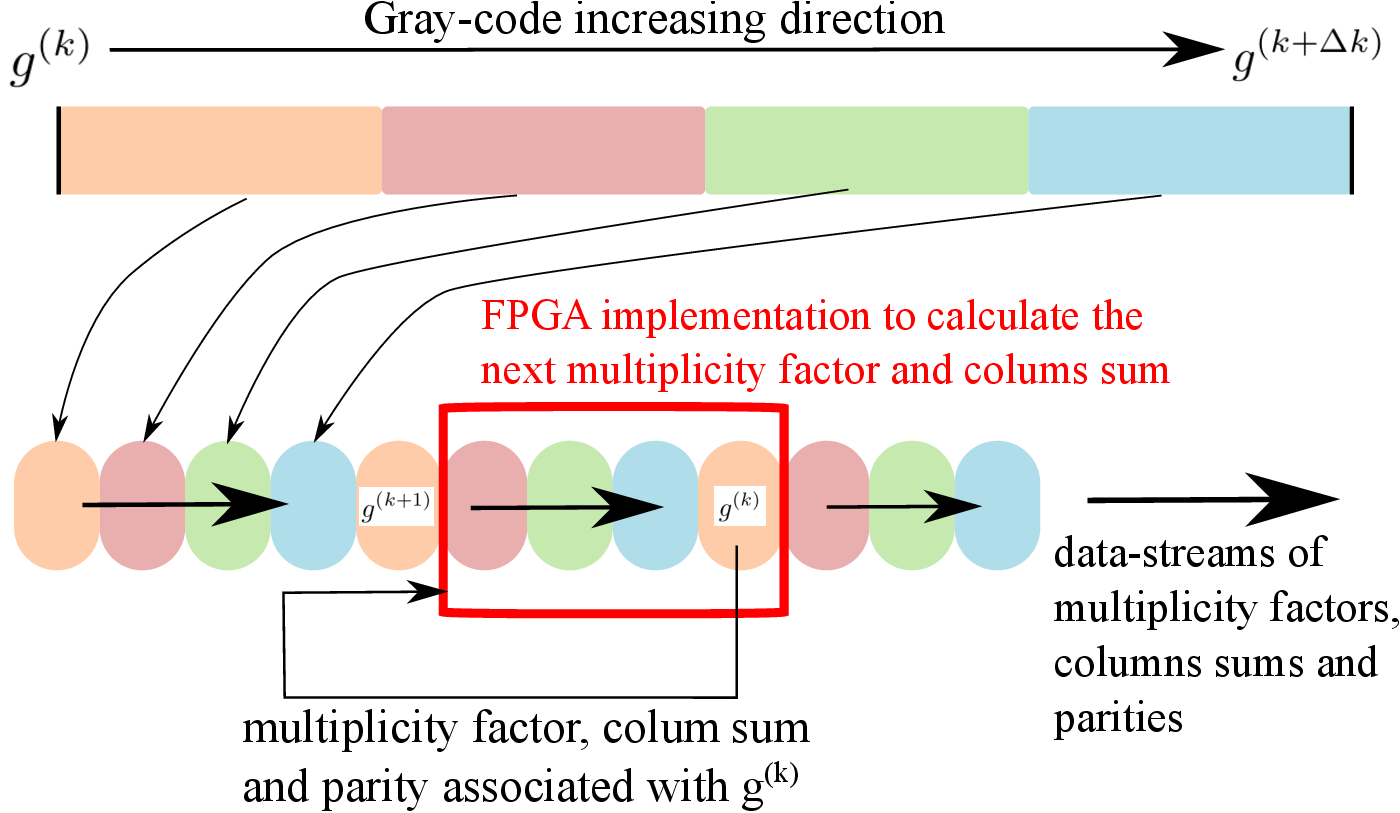}
    \caption{The staggered strategy to calculate the multiplicity (binomial) coefficients during permanent evaluation with repeated rows and columns.
    The coefficients associated with the individual Gray codes are calculated recursively from the value associated with the preceding Gray code. 
    In the Figure, we assume that the calculation of the new multiplicity factor takes $L=4$ ticks (in our actual DFE implementation the latency of the cycle is at least $L=9$), so $4$ Gray-code streams are alternating streamed in the column sum kernels (green colored kernels in Fig.~\ref{fig:DFE_design}.) to ensure smooth data flow design providing an addend to the permanent in each tick.
    Further details of the scheme are discussed in Sec.~\ref{sec:repeatedDFE}.}
    \label{fig:staggered_gray_code}
\end{figure}

Unfortunately, the calculation of the multiplicity factors cannot be calculated in one tick (like the column sums in the green-colored kernels of Fig.~\ref{fig:DFE_design}), and hence they cannot be reused in the next tick to calculate the forthcoming multiplicity factor, breaking up the "smooth" data-flow nature of our implementation calculating an addend to the permanent in each tick.
To overcome this issue, we split up the range of the Gray codes processed by a single column sum kernel into $L$ chunks, where $L$ is bigger or equal to the number of clock cycles needed to update a multiplicity factor. 
We use loop-tiling to pipeline the Gray-code chunks in a way that the "distance in time" between Gray codes coming from the same chunk is $L$ cycles.
(The idea is demonstrated in Fig.~\ref{fig:staggered_gray_code} for the example of $L=4$.)
This way we end up with $L$ staggered Gray code counters embedded into a single stream of a counter.
The column sum kernels of Fig.~\ref{fig:DFE_design} are adapted to this staggered nature of the counter and each column sum is reused to provide the next column sum in the forthcoming $L$-th cycle, while the other $L-1$ columns sums and multiplicity factors are buffered on the chip waiting to be reused for data-stream generation. 
In the meantime, the copies of the calculated column sums and multiplicity factors are streamed towards product kernels in a way described in Sec.\ref{sec:DFE} providing data for addends to the permanent in each tick.

Finally, we would like to mention, that in order to split the calculation between $4$ (or $8$ kernel is dual mode) one needs to introduce $2$ ($4$) rows with a multiplicity of unity (besides the very first row). 
If there are no sufficient single-multiplicity rows in the matrix, we extract the necessary number of rows from nontrivial-multiplicity rows and treat them as single-multiplicity rows.
By incorporating the outlined additional logic on the FPGA chip we managed to compile the repeated row DFE implementation for matrices having at most $40$ columns, working at clock frequency $330$ MHz.

\subsection{Collecting the input matrices into batches} \label{sec:batch}

As our numerical results demonstrated in Sec.~\ref{sec:numerical_results}, the PCIe communication overhead dominates the execution time at smaller matrices.
However, the DFE can be configured to stream and process multiple matrices (arranged into a continuous memory block on the CPU side) in a single execution. 
\begin{figure}
    \centering
    \includegraphics[width=0.6\textwidth]{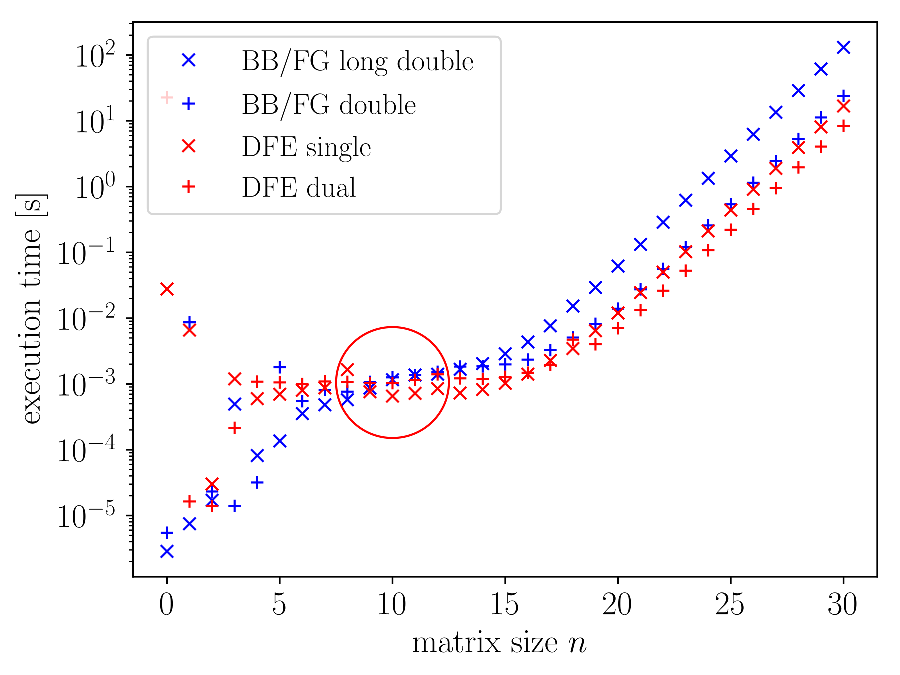}
    \caption{Batched performance comparison of the permanent calculator DFE to the BB/FG implementations of the Piquasso Boost library for CPU.
    During the benchmark, the processed random unitaries (for which the permanent implementations were evaluated) had no repeated rows and columns. 
    The CPU measurements were done by averaging the execution time of $30$ computations.
    For single and dual DFE the $30$ matrices were uploaded and processed in one shot, i.e. in batched mode.
    The PCIe overhead is split between the $30$ matrices, significantly reducing the CPU-DFE crossover.    
    }
    \label{fig:batched_benchmark}
\end{figure}
This way the PCIe overhead becomes insignificant compared to the overall computational time, reducing the CPU-DFE crossover to lower matrix sizes than it was reported in Figs.~\ref{fig:benchmark} and \ref{fig:benchmark_repeated} in the non-batched execution. 
\begin{figure}
    \centering
    \includegraphics[width=0.9\textwidth]{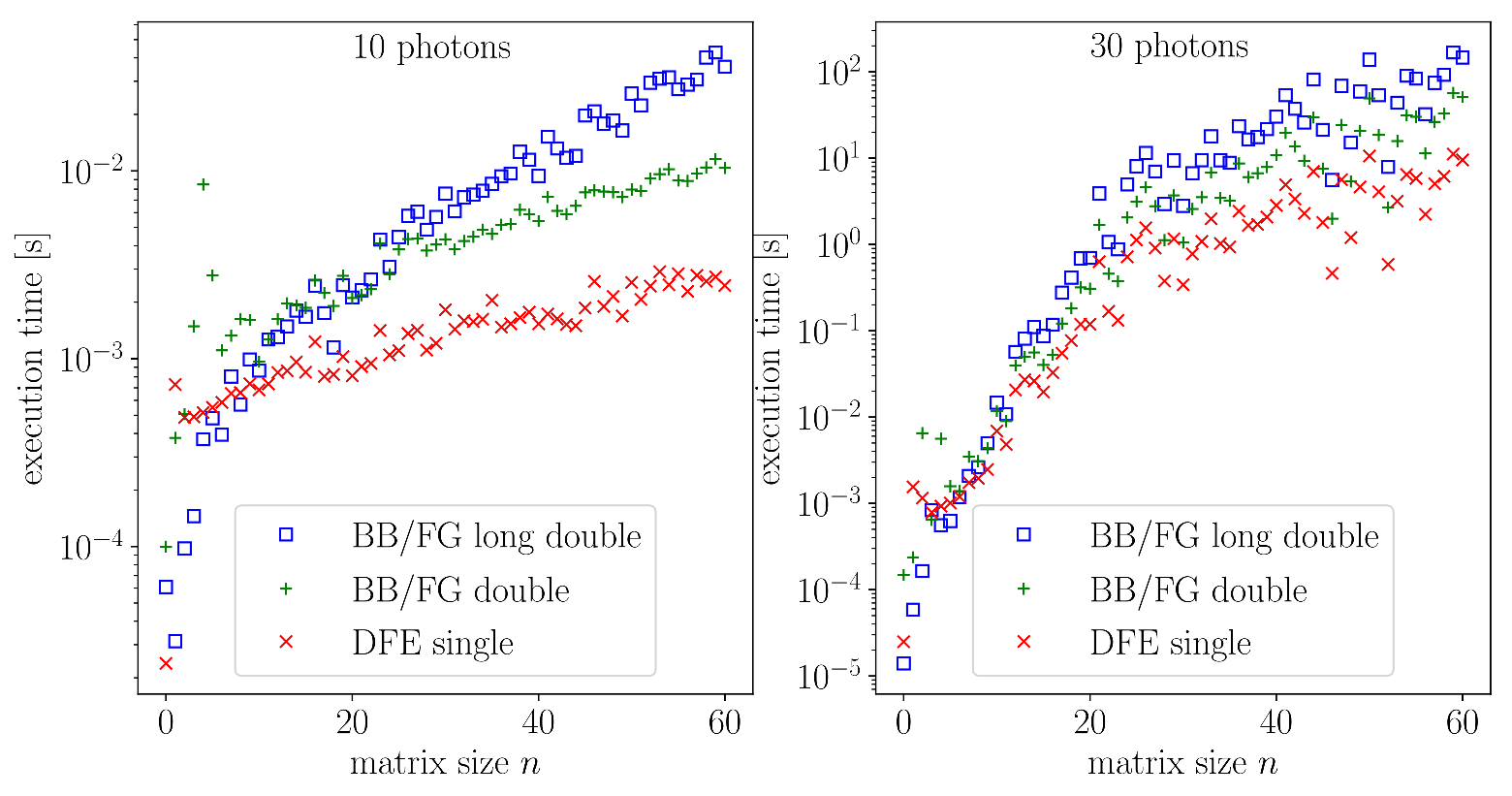}
    \caption{Batched performance comparison of the permanent calculator DFE to the BB/FG implementations of the Piquasso Boost library for CPU accounting for row and col multiplicities in the input matrix.
    During the benchmark, the processed unitaries (for which the permanent implementations were evaluated) and row/column multiplicities were selected randomly adding up to $10$ (left) and $30$ (right) photons. 
    The number of the batched matrices processed in a single execution on the DFE was chosen to be equal to the matrix size $n$.
    The plotted execution time stands for the total execution time divided by the number of the processes matrix instances.
    The PCIe overhead is split between the processed matrices significantly reducing the CPU-DFE crossover.  
    }
    \label{fig:batched_benchmark_repeated}
\end{figure}
Fig.~\ref{fig:batched_benchmark} shows our batched DFE performance results compared to the CPU implementations.
In this specific case, $30$ matrices were processed through the DFE for each matrix size $n$ in a single, batched execution.
The CPU-DFE crossover has now been reduced to a matrix size of $n\sim10$.
In order to support batched execution mode, additional logic has been implemented on the DFE to reset all the necessary counters within the DFE computing kernels each time the prescribed clock cycles to calculate new permanent elapses.
For the sake of simplicity, our implementation requires that the batched matrices must be of equal size.
This way only a few scalar values need to be streamed up to the FPGA, leaving the optimized DFE implementation mostly unchanged, enabling us to compile the implementation for high frequencies already achieved in Sec.~\ref{sec:DFE}.
Similar logic can be applied to the DFE implementation including row and column multiplicities.
Figure \ref{fig:batched_benchmark_repeated} shows our performance benchmark on batched permanent evaluation accounting for multiple occupancy of photonic modes.
The execution time plotted in these figures stands for the batched permanent evaluation time averaged over the number of matrices in the batch, which was chosen to be equal to the matrix size describing the photonic mode count.
The input unitaries were chosen randomly together with the row and column multiplicities. 
Similarly to the non-repeated variant of the permanent calculator DFE, the execution time of the individual permanent evaluations should be equal to be able to systematically reset the counters while keeping up a steady data flow. 
This can be achieved by batching up permanent calculation problems having identical row multiplicities, but different column selections.
This requirement might seem to be too restrictive, but it fits the boson sampling algorithm of \cite{clifford2020faster} very well, in which many permanents need to be evaluated with different input states (column selections), but identical output states (row selections).

As one can see, the advantage of DFE over CPU implementations is evident even at photon counts as low as $\sim10$ providing a significant DFE speedup in the boson sampling simulation discussed in the main text.
While the described batching strategy can be further generalized, it already fits to many realistic use cases in photonic quantum computations.
We leave further generalization of the implementation for future work.

\subsection{Normalization of the input matrices processed through the DFE} \label{sec:normalization}

In this section, we provide numerical details on preventing fixed point number representation overflow during arithmetic operations and keep up the numerical accuracy of the permanent evaluation.
We achieve the numerical stability by scaling the input matrix on the CPU side, evaluating the permanent function on the DFE for the modified matrix, and re-scale the retrieved value from the DFE to get the correct result for the permanent. 
The scaling of the input matrix is performed column-wise by multiplying the elements in a column with the same factor but using different factors in each of the columns. 
According to the BB/FG formula (\ref{eq:bbfg}) and (\ref{eq:bbfgrepeated}) with row and column multiplicities the DFE result then needs to be re-scaled by the product of the scaling factors $\alpha_j$, by taking the multiplicity of the individual factors corresponding to the column multiplicities in the input matrix.   
Now we briefly describe our mathematical considerations when determining the normalization factors. 
In order to prevent overflow during the calculation carried out by fixed point number representation (keeping the number of integer and fractional bits constant), we bound the inner column sum of the scaled input matrix (with complex values) such that
\begin{equation}
    \textrm{max}\left(\left|\sum\limits_{i=0}^n \delta_i \frac{a_{i,j}}{\alpha_j}\right|\right) \leq 1\; \forall j,
\end{equation}
allowing for a compact fixed-point representation with a sign bit, a bit for integral value and the remaining for decimal places.  
In order to determine the scaling factors, for each column $j$ we look at the problem geometrically with the complex numbers interpreted as Euclidean vectors in a two-dimensional space.
To identify the worst case (without processing all the variations of the $\boldsymbol{\delta}$ vectors individually), first we sum up the complex numbers $\pm a_{i,j}$ corresponding to the four quadrants of the complex plane independently according to the sign of the real and imaginary parts of $a_{i,j}$. This step results in four distinct vectors, obtained by summing up the matrix elements in the individual quadrants with values $\delta_i=1$ $\forall i$.
Then we subtract the resulting vectors in quadrant pairs ($I$,$III$) and ($II$,$IV$) leading to two distinct vectors. 
This step introduces the possible $\delta_i=-1$ values into the summation. 
Finally, we compare the addition and subtraction of the two vectors to end up with a final vector of the largest possible magnitude determining the scaling coefficient $\alpha_j$. 
The complexity of this normalization procedure is $\mathcal{O}(n^2)$ being negligible compared to the evaluation complexity of the permanent function.

The final point is that the outer sum of the BB/FG formula will always result in a normalized value, but the partial sums may not.  This is dependent on the order of summations and is typically exacerbated when binomial coefficient methods are used.  Therefore, the computation here is simpler as the real and imaginary parts can be treated separately.  We consider an all-one matrix as a worst case, which when normalized has all values $\frac{1}{n}$. Therefore by looking at the worst-case positive and negative partial sums based on
\begin{equation}
    \frac{1}{n^n}\max\left(\sum\limits_{k=0}^{\lfloor(n-1)/2 \rfloor} (2(2k+2)-n)^n {n-1 \choose 2k+1}, \sum\limits_{k=0}^{\lfloor(n-1)/2 \rfloor} (2(2k+1)-n)^n {n-1 \choose 2k} \right)
\end{equation}
e.g. for $n=40$, then $~\pm 27$ is the maximum partial sum requiring $6$ integer bits (including sign).

\end{document}